\documentclass[11pt]{article}
\usepackage[utf8]{inputenc}
\usepackage{amsmath, amssymb, enumerate, graphicx, fullpage, physics}

\usepackage{sidecap}
\usepackage{titling}
\usepackage{setspace}
\usepackage[margin=1in]{geometry}
\usepackage{overpic, rotating, pbox}

\usepackage{palatino} 

\newcommand{\R}{\mathbb{R}}
\newcommand{\St}{\mathrm{St}}
\newcommand{\kB}{k_\mathrm{B}}

\usepackage[dvipsnames]{xcolor}
\definecolor{mpl_orange}{RGB}{255, 127, 14}

\title{  Nonlinear stochastic modeling with Langevin regression
	\vspace{-0.5cm}}
\date{}
\author{Jared L. Callaham, Jean-Christophe Loiseau, Georgios Rigas, and Steven L. Brunton}

\begin{document}

\maketitle
\vspace{-1.5cm}

\begin{abstract}
	Many physical systems characterized by nonlinear multiscale interactions can be effectively modeled by treating unresolved degrees of freedom as random fluctuations.
	However, even when the microscopic governing equations and qualitative macroscopic behavior are known, it is often difficult to derive a stochastic model that is consistent with observations.
	This is especially true for systems such as turbulence where the perturbations do not behave like Gaussian white noise, introducing non-Markovian behavior to the dynamics.
	We address these challenges with a framework for identifying interpretable stochastic nonlinear dynamics from experimental data, using both forward and adjoint Fokker-Planck equations to enforce statistical consistency.
	If the form of the Langevin equation is unknown, a simple sparsifying procedure can provide an appropriate functional form.
	We demonstrate that this method can effectively learn stochastic models in two artificial examples: recovering a nonlinear Langevin equation forced by colored noise and approximating the second-order dynamics of a particle in a double-well potential with the corresponding first-order bifurcation normal form.
	Finally, we apply the proposed method to experimental measurements of a turbulent bluff body wake and show that the statistical behavior of the center of pressure can be described by the dynamics of the corresponding laminar flow driven by nonlinear state-dependent noise.
\end{abstract}

\section{Introduction}

It is widely accepted in physics that nominally deterministic systems with many degrees of freedom can often be modeled more effectively from a statistical point of view.
In many complex multiscale systems, a variety of processes lead to emergent large-scale structures whose dynamics are described by a relatively small set of macroscopic variables~\cite{Haken1983, Cross1993}.
The influence of the unresolved degrees of freedom can be approximated with random forcing in the spirit of statistical mechanics~\cite{Risken1996book, Friedrich2011review, RezaRahmiTabar2019book}.
This stochastic treatment of unresolved variables has become commonplace in fields including climate science~\cite{Majda2006book}, ecology~\cite{Levin1992}, epidemiology~\cite{Allen2008epidemiology}, protein folding~\cite{Prinz2011jcp}, neuroscience~\cite{Harrison2005neuro}, and turbulence~\cite{Kraichnan1989}.

\begin{figure}
	\centering
	\vspace{.1in}
	\begin{overpic}[width=1\linewidth]{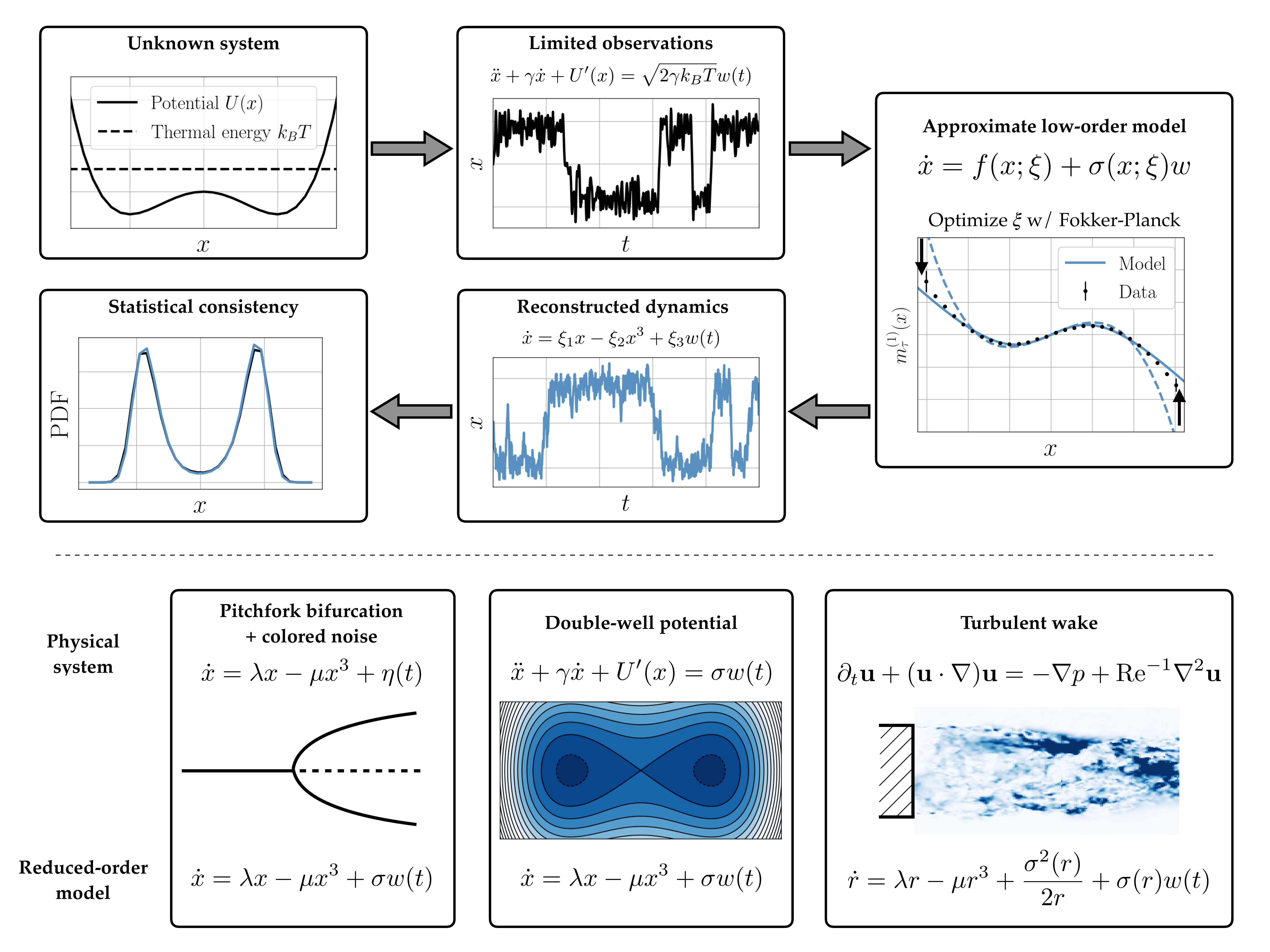}
	\end{overpic}
	\vspace{-.3in}
	\caption{ \small{\textbf{Schematic of Langevin regression (top) with example applications (bottom).}
			Given a long time series of a macroscopic variable describing a complex system, we seek to identify an approximate stochastic model.
			The variable $x$ might represent a reaction coordinate capturing metastable protein configurations or the temporal coefficient of a dominant global hydrodynamic mode, for instance.
			Langevin regression uses both the forward and adjoint Fokker-Planck operators to optimize free parameters $\xi$ of the model, ensuring consistency with observed statistics such as the finite-time Kramers-Moyal coefficients $m_\tau^{(n)}(x)$ (see Sec.~\ref{sec:theory}).
			}
	}
	\label{fig:overview}
\end{figure}

The stochastic evolution of a state $x$ are often represented with Langevin dynamics
\begin{equation} \label{eq:langevin}
\dot{x} = f(x) + \sigma(x) w(t).
\end{equation}
The deterministic ``drift" dynamics $f(x)$ describe the evolution of the slow macroscopic variables, while the fluctuations are parameterized by the diffusion term $\sigma(x) w(t)$, where $w(t)$ is typically assumed to be a Gaussian white noise process.
If this model cannot be derived from first principles, a model can sometimes be inferred from observations of the natural dynamics of the system, as illustrated in Fig.~\ref{fig:overview}

A significant challenge in constructing approximate stochastic models is that many of these systems are far enough from thermal equilibrium that even when the microscopic governing equations are known, fundamental principles such as detailed balance and the fluctuation-dissipation theorem cannot be readily applied.
For example, widely separated time scales for forcing and dissipation prevents viscous turbulence from approaching a state of equipartition~\cite{Kraichnan1989, Noack2008jnet}.
This scale separation is captured by the Reynolds number, which can be interpreted as a ratio of the time scales characterizing the energetic large-scale dynamics and the small-scale viscous motions~\cite{Tennekes1972book}.

As with equilibrium statistical physics and quantum mechanics, there are several ways to represent stochastic dynamics, as exemplified by the differing treatments of Brownian motion by Einstein~\cite{Einstein1905brownian} and Langevin~\cite{Langevin1908}.
Einstein's theory is constructed around a diffusion equation governing the evolution of the distribution of particles, while Langevin's describes an individual trajectory of a particle subject to friction and a random fluctuating force.
This duality persists in the modern theory; the same stochastic process can be represented with a generalized Langevin-type differential equation governing trajectories or a Fokker-Planck equation for the evolution of the probability distribution~\cite{Risken1996book}.

Linear dynamics can be analyzed from either perspective, but in the nonlinear case the most convenient representation often depends on the application.
Nonlinear Langevin-type stochastic differential equations are difficult to treat analytically, but fit more naturally with low-dimensional modeling and control objectives.
On the other hand, Fokker-Planck equations replace nonlinear trajectory dynamics with a linear partial differential equation for the probability distribution.
The ensemble perspective also facilitates comparison with long time series measurements of ergodic systems.

In this work we seek to exploit these equivalent representations to identify Langevin dynamics by using both the forward and adjoint Fokker-Planck equations to ensure consistency with observations.
We propose a framework for identifying nonlinear stochastic models from noisy experimental data, building on previous work in sparse regression~\cite{Brunton2016pnas, Boninsegna2018jcp} and adjoint-based parameter estimation~\cite{Honisch2011pre}.
After presenting background material on stochastic dynamics in Sec.~\ref{sec:theory}, we describe the method in Sec.~\ref{sec:method}.
Finally, we demonstrate its application to three example systems in Sec.~\ref{sec:results}: a cubic model driven by colored noise, a particle in a double-well potential, and a symmetry-breaking instability in a turbulent wake.

\subsection{Related work}
As we study increasingly complex systems that deviate from the restrictive near-equilibrium conditions of classical statistical mechanics, fully empirical system identification becomes more appealing.
The primary goal of these methods is not to simply fit the statistics, but to identify a minimum-complexity mechanistic model that describes the important interactions in the system~\cite{Billings2013book}.
The resulting models stand in for those derived via traditional analysis and can be used to gain physical insight into the system.
Secondary goals in engineering fields can include the design of control systems~\cite{Brunton2015amr} or parametric surrogate models for design and optimization~\cite{Benner2015siam}.
Predicting response to exogenous inputs or parametric variation is clearly more difficult, but a dynamic model of the autonomous behavior is a necessary prerequisite to these aims.


Stochastic systems can be broadly categorized according to the type of dynamics, noise, and stochasticity.
Dynamics may be linear or nonlinear, the noise process may be white (uncorrelated in time) or colored (time-correlated), and the strength of the fluctuations may be constant (additive diffusion) or state-dependent (multiplicative diffusion).
Similarly, stochastic model identification methods can be similarly categorized by the type of models they are able to construct.

For example, realization algorithms are a mainstay of engineering disciplines, although these methods are restricted to linear input/output systems with additive white noise~\cite{Juang1985, Juang1991, VanOverschee1994}.
Perhaps the most general and successful nonlinear approach is the NARMAX framework, which can construct nonlinear models driven by state-dependent colored noise~\cite{Billings2013book}.
However, NARMAX models typically cannot be transformed to continuous time, which is often the most natural setting for physical problems, making them difficult to interpret.
More recent work has explored a variety of strategies for modeling nonlinear stochastic systems, including operator theoretic methods~\cite{Klus2020physica}, optimal transport~\cite{Arbabi2019}, deep learning~\cite{Wang2019acs, Noe2019science, Noe2020annrev}, and identifying distribution evolution equations~\cite{Bakarji2020}, although none of these pursues a representation in terms of nonlinear state-space dynamics.
On the other hand, recent work has demonstrated that a stable linear system driven by colored noise can accurately reproduce second-order turbulent statistics~\cite{Zare2017jfm}.

Recent advances have made significant inroads towards continuous-time model discovery in deterministic nonlinear systems~\cite{Kantz2003book, Schmidt2009science}.
For example, sparse identification of nonlinear dynamics (SINDy) approximates time derivatives with a sparse linear combination of candidate functions~\cite{Brunton2016pnas}.
However, even without the difficulties of estimating time derivatives from noisy data, a major challenge for extending deterministic methods to stochastic modeling lies in disambiguating the macroscopic dynamics from the unresolved degrees of freedom.
In cases where the dynamics can be closely approximated by one-dimensional dynamics forced by additive white noise, parameters may be identified by regression to analytic solutions of the PDF~\cite{Rigas2015jfm, Brackston2016jfm}. 
For more general systems, recent work has approached the parameter estimation problem with inference methods based on ensemble Kalman filtering~\cite{Schneider2020} and information theory~\cite{Frishman2020prx, Bruckner2020prl}.
Alternatively, Boninsegna \textit{et al.} introduced a major contribution to stochastic modeling by demonstrating that SINDy could be extended to stochastic systems without Monte Carlo approximation via the conditional moments used in the Kramers-Moyal expansion~\cite{Boninsegna2018jcp}.
This stochastic SINDy method was capable of recovering the correct model structure and parameters from large libraries of candidate functions.

Approximating stochastic dynamics from data with the Kramers-Moyal average has a long history of successful modeling in a wide range of fields~\cite{Friedrich2011review}.
However, as with many theoretical results, it is predicated on the assumption that the dynamics are driven by Gaussian white noise. 
As recognized by Einstein, even the molecular forcing involved in Brownian motion has some finite decorrelation time since it is a continuous physical system~\cite{Einstein1905brownian}.
Moreover, omitting degrees of freedom from an otherwise Markovian\footnote{In this context, meaning that the evolution of the system only depends on its current state and not its time history} system generally leads to explicit memory effects in the dynamics~\cite{Zwanzig2001}.
We therefore expect that all systems will have some characteristic ``Einstein-Markov" time scale over which the time evolution of macroscopic variables may depart significantly from the standard assumptions of stochastic modeling~\cite{Friedrich2011review}.
For Brownian motion this time scale is on the order of picoseconds, while for complex, multiscale, far-from-equilibrium systems it may even be longer than experimental sampling rates.

\begin{SCfigure}
	\centering
	\begin{overpic}[width=0.6\linewidth]{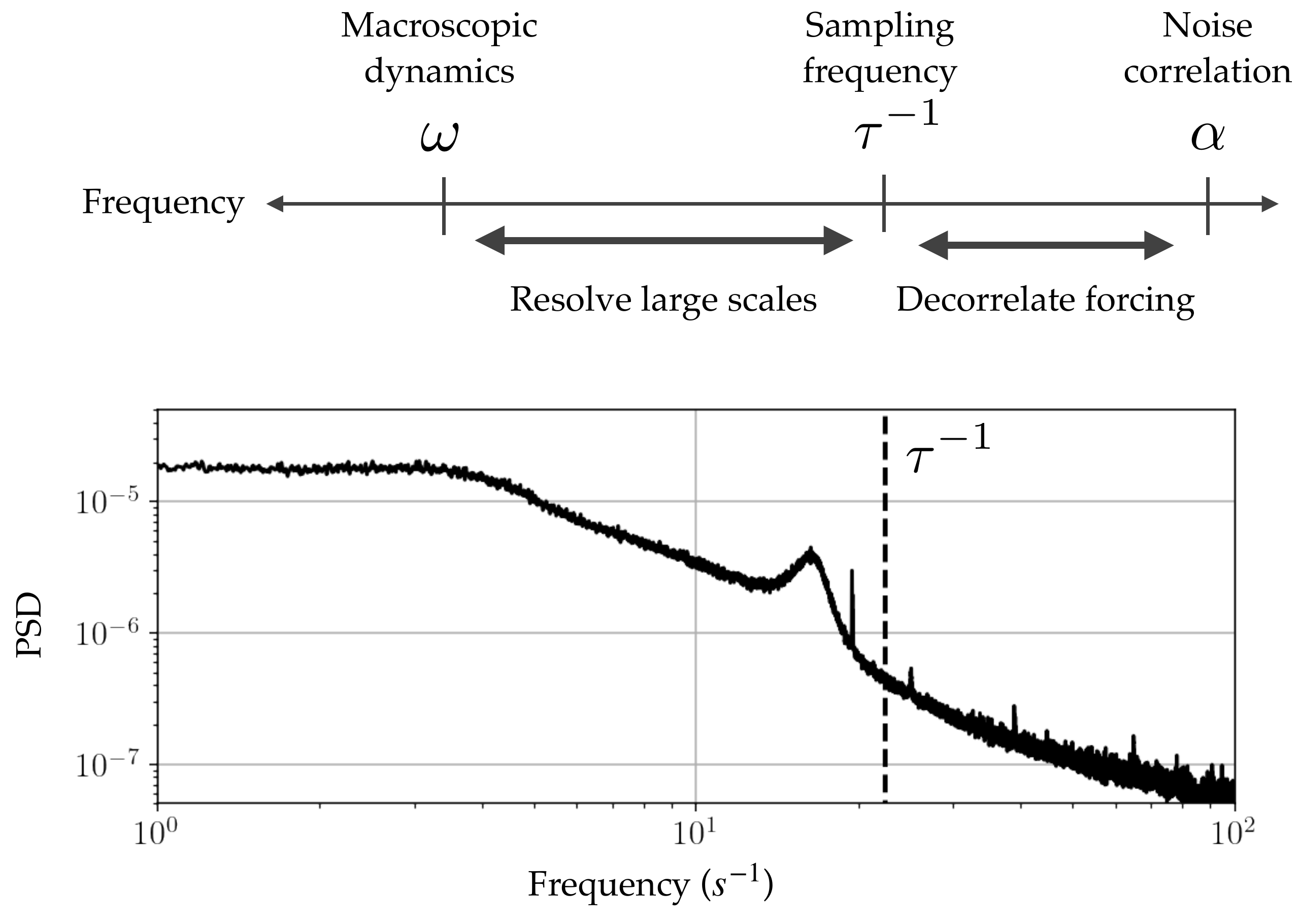}
	\end{overpic}
	\vspace{-.1in}
	\caption{ Dual scale separation for stochastic modeling.
		Even the fastest scales of continuous physical systems are characterized by some finite decorrelation rate $\alpha$ (e.g.~\eqref{eq:colored}).
		However, if the macroscopic dynamics have a much slower characteristic timescale $\omega \ll \alpha$, we may be able to choose a sampling rate $\tau^{-1}$ which can simultaneously resolve the dominant dynamics and allow the unresolved scales to decorrelate.
		For example, the power spectrum of the radial center of pressure of the turbulent wake is shown at bottom along with the subsampling rate used in Sec.~\ref{sec:results-wake}.
	}
	\label{fig:scale-separation}
\end{SCfigure}

These considerations make the sampling rate used to construct stochastic models from experimental time series an important choice.
Theoretical difficulties introduced by time-correlated forcing and non-Markovian effects can be avoided to some extent by deliberately subsampling.
This has been established in finance~\cite{Zhang2005finance, AitSahalia2005finance} and shown for artificial dynamics with two widely separated time scales~\cite{Pavliotis2007}.
Qualitatively, coarse sampling allows the unresolved degrees of freedom to decorrelate, while ideally still resolving the coherent macroscopic scales (see Fig.~\ref{fig:scale-separation}).
If the fluctuations appear uncorrelated in time, standard theoretical tools, such as the Kramers-Moyal average, may once again be applied.
However, coarse sampling leads to distorted estimates of the conditional moments used in the Kramers-Moyal expansion~\cite{Ragwitz2001prl}, although these finite-time effects can be accounted for using the adjoint Fokker-Planck equation~\cite{Lade2009}.

The relevance of this result for parameter estimation in stochastic models was realized by Honisch \& Friedrich, who proposed an optimization framework designed to correct for finite sampling-rate effects~\cite{Honisch2011pre}.
This technique has recently been refined and applied to parameter estimation for amplitude equations describing several different physical systems by Boujo, \textit{et al}~\cite{Boujo2016prs, Boujo2019jfs, Boujo2020jsv}.
As with the majority of work in nonlinear time series analysis~\cite{Takens1981, Kantz2003book, Bradley2015chaos}, existing studies have focused on modeling scalar observables, although the theory readily generalizes to complex- or vector-valued systems.

\subsection{Contributions}
Here we show that these finite-time corrections generalize the stochastic SINDy method~\cite{Boninsegna2018jcp} to the broad class of systems for which the forcing cannot be treated as white noise.
Specifically, we explore systems for which the fast scales have nontrivial dynamics, the exclusion of which formally breaks the Markovian properties of the full physical system.
This includes colored noise, latent variables, and ``microscopic" degrees of freedom with significantly nonzero time correlations.
Furthermore, we extend the adjoint Fokker-Planck optimization problem with the forward steady-state solution to enforce consistency between the model and the empirical probability distribution. 

The proposed modeling framework, which we refer to as Langevin regression, is designed to identify nonlinear Langevin-type equations directly from noisy experimental data.
This method combines the advantages of three previously distinct approaches: adjoint-based parameter estimation with the Kramers-Moyal average~\cite{Lade2009, Honisch2011pre}, learning unknown model structure with sparse regression~\cite{Brunton2016pnas, Boninsegna2018jcp}, and steady-state PDF fitting~\cite{Rigas2015jfm}.
Central aspects of our approach have been introduced in these previous works,  but here we develop a single, unified framework.
The main contributions of this work are detailed in Sec.~\ref{sec:method} and may be summarized as follows:
\begin{enumerate}
	\item Unresolved degrees of freedom often have nontrivial time correlations.
	We show that with deliberate subsampling and finite-time corrections, macroscopic variables can be modeled by low-order nonlinear dynamics driven by uncorrelated white noise.
	\item Finite-time effects due to coarse sampling rates can be corrected with parameter estimation based on the adjoint Fokker-Planck equation.
	We extend this optimization with the forward solution, enforcing consistency with the steady-state probability distribution.
	\item If the form of the stochastic model is unknown, its structure can be automatically identified with an iterative SINDy-type model selection procedure.
	We show that this stepwise sparse regression can be straightforwardly combined with the Fokker-Planck optimization.
\end{enumerate}

We explore the proposed nonlinear stochastic model identification method on several example systems.
First, we illustrate the importance of judicious subsampling and finite-time corrections for correlated forcing by recovering a nonlinear Langevin equation driven by colored noise.
We then show that the Langevin regression can construct a  reduced-order model approximating the second-order dynamics of a particle in a double-well potential with a first-order bifurcation normal form.
Both of these illustrative examples avoid the latent variable problem for the unresolved degrees of freedom by learning stochastic closure models.
An implementation of Langevin regression along with code to reproduce the results from the simulated system is available on GitHub\footnote{https://github.com/dynamicslab/langevin-regression}.

Finally, we apply Langevin regression to experimental measurements of a turbulent bluff-body wake; the sparse model selection procedure identifies a model similar to that proposed by Rigas \textit{et al}~\cite{Rigas2015jfm}, but with an additional nonlinear noise term that improves the correspondence with both the empirical probability distribution and power spectral density.
Langevin regression draws from both the long legacy of stochastic modeling and recent advances in data-driven methods to form a flexible and general framework for approximating complex nonlinear dynamics with statistically consistent stochastic models.


\section{Background on stochastic dynamics}
\label{sec:theory}

This section briefly reviews relevant theoretical concepts in stochastic modeling, including the Fokker-Planck equation, the Kramers-Moyal conditional average, and adjoint corrections for finite-time sampling effects.
For more comprehensive background on the topics of nonequilibrium statistical mechanics and the physical applications of stochastic differential equations we refer the reader to excellent reference texts such as Refs.~\cite{Risken1996book, Zwanzig2001, RezaRahmiTabar2019book}.

In this work we seek to model the macroscopic variables $x \in \R^d$ with Langevin dynamics of the form given by Eq.~\eqref{eq:langevin}.
The influence of unresolved degrees of freedom on the standard deterministic dynamics $\dot{x} = f(x)$ is modeled with the diffusion term $ \sigma(x) w(t) $, where $w(t)$ is a white noise process.
The diffusion is called additive if $\sigma(x)$ is a constant, or multiplicative if it is state-dependent.
The majority of stochastic models assume additive noise, since is easier to treat analytically and is often a reasonable approximation for systems without strong coupling across scales.
For instance, additive process noise is the standard assumption for linear state-space models of electronic circuits subject to thermal fluctuations.
However, for systems such as fluid dynamics where the quadratic nonlinearity leads to bidirectional coupling between the coherent and turbulent degrees of freedom, state-dependent noise may improve the model~\cite{Majda2001cpam}.

Perhaps the most restrictive assumption is that placed on the noise process $w(t)$.
In order to treat Langevin dynamics analytically, the forcing is typically taken to be Gaussian-distributed and delta-correlated in time, i.e. $\langle w(t) w(t') \rangle = \delta(t - t')$.
If this is not the case, the following discussion becomes more complicated~\cite{Risken1996book, Zwanzig2001}.
For a macroscopic or integral quantity, the assumption of Gaussian statistics can be argued by appealing to the central limit theorem, but the time-correlation requirement is generally more difficult to justify.

\subsection{Fokker-Planck equation}
The Langevin equation~\eqref{eq:langevin} describes individual trajectories, but due to the variability inherent in stochastic dynamics, it is often more natural to approach stochastic dynamics from the ensemble perspective.
For the Langevin dynamics given by Eq.~\eqref{eq:langevin}, conservation of probability requires that the probability density function (PDF) $p(x,t)$ evolves in time according to the \textit{Fokker-Planck equation}\footnote{Here and throughout we use the It\^o interpretation of stochastic integrals.}:
\begin{equation}
\label{eq:forward-fp}
\pdv{p(x,t)}{t} = -\pdv{x_i} \left[ f_i (x) p(x, t) \right] +\frac{\partial^2}{\partial x_i \partial x_j}  \left[ a_{ij}(x) p(x, t) \right] \equiv \mathcal{L} p,
\end{equation}
where the diffusion tensor $a(x)$ is given by $a_{ij}(x) =  \sigma_i(x)\sigma_j(x)/2$ and the subscripts indicate Einstein summation notation.
As an alternative approach to Langevin equations, the multiscale interactions in complex spatiotemporal systems can be modeled explicitly with Fokker-Planck equations~\cite{Peinke2019annrev}.

Often we are interested in the case where the system is statistically stationary, so that $\mathcal{L}p = 0$ and $p = p(x)$.
For example, if $x$ is a scalar and the diffusion $\sigma$ is a constant, the steady-state solution can be determined analytically:
\begin{equation}
\label{eq:1d-ss-fp}
p(x) = C \exp \left[ \frac{2}{\sigma^2} \int f(x) \dd x \right],
\end{equation}
where the constant $C$ is determined by the normalization condition $\int p(x) \dd x = 1$.
However, solving the Fokker-Planck equation for general nonlinear dynamics is challenging and typically must be approached approximately or numerically~\cite{Risken1996book}.

\subsection{Kramers-Moyal average}
The Fokker-Planck equation may also be derived by expressing the time evolution of a general PDF as a Taylor series of conditional finite-time moments $m_\tau^{(n)} (x)$,
where the conditional finite-time moments $m_\tau^{(n)} (x)$ are
\begin{equation} \label{eq:km-avg-definition}
m_\tau^{(n)} (x) = \left\langle \left( x'(t+\tau) - x'(t) \right)^n \right \rangle_{x'(t) = x}.
\end{equation}
This leads to the \textit{Kramers-Moyal expansion}.
For scalar $x$,
\begin{equation}
\pdv{p}{t} = \lim_{\tau \rightarrow 0} \frac{1}{\tau} \sum_{n=1}^\infty \left( -\pdv{x} \right)^n \frac{m_\tau^{(n)}(x)}{n!} p(x, t).
\end{equation}
Viewing the conditional mean in Eq.~\eqref{eq:km-avg-definition} as a finite-difference formula, the first moment $m_\tau^{(1)}(x)$ gives the average displacement over a time interval $\tau$ if the system is in state $x$.
Likewise, the second moment $m_\tau^{(2)}(x)$ gives a conditional variance of the short-time evolution.

However, according to Pawula's theorem, if the system is driven by Gaussian white noise all moments $ n \geq 3$ vanish~\cite{Risken1996book}, leading to the Fokker-Planck equation~\eqref{eq:forward-fp}.
The leading moments are related to the drift and diffusion functions of the corresponding Langevin equation in the limit of vanishing time interval, i.e.
\begin{subequations}
	\begin{align}
	\label{eq:km-drift}
	f_i(x) &= \lim_{\tau \rightarrow 0} \frac{1}{\tau}  \left\langle x'_i(t + \tau) - x'_i \right\rangle_{x'(t) = x }\\
	\label{eq:km-diffusion}
	a_{ij}(x) &= \lim_{\tau \rightarrow 0} \frac{1}{2\tau}   \left\langle (x_i(t + \tau) - x_i) (x_j(t + \tau) - x_j)  \right\rangle_{x'(t) = x }.
	\end{align}
\end{subequations}
We refer to these relationships as the \textit{Kramers-Moyal average}.

\begin{figure}
	\centering
	\vspace{.3in}
	\begin{overpic}[width=0.95\linewidth]{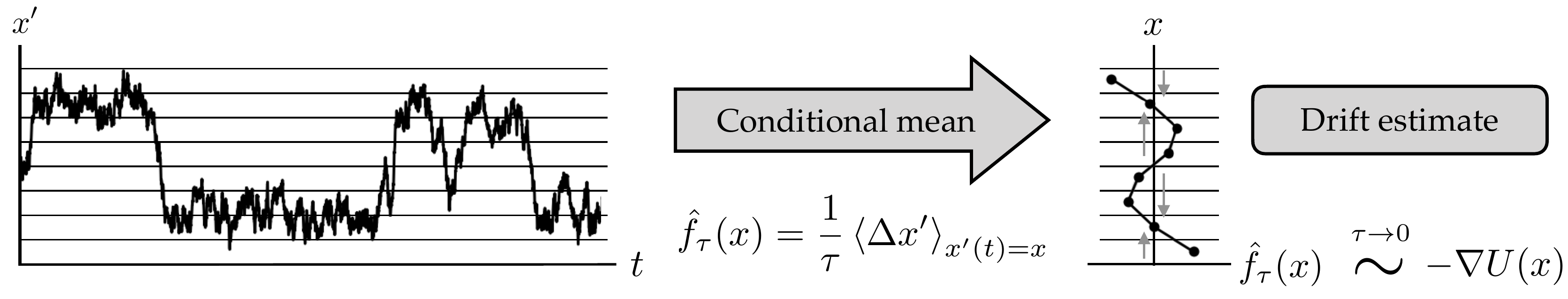}
	\end{overpic}
	\vspace{-.1in}
	\caption{ Schematic of Kramers-Moyal coefficient estimation for the first moment (drift). The drift estimate is determined by the conditional mean of the state evolution over the interval, while the diffusion is given by the conditional variance.
		The conditional moments can be approximated by dividing a long time series into histogram bins and taking the mean and variance within each bin.
		For example, the Kramers-Moyal drift estimate gives an approximate discretized vector field for the deterministic component of the dynamics (right).
	}
	\label{fig:km-schematic}
\end{figure}

In principle, these averages could offer a way to approximate unknown drift and diffusion functions from data by binning the time series into histograms and computing~\eqref{eq:km-drift} and~\eqref{eq:km-diffusion} with the sampling rate $\tau = 1/f_s$, as illustrated in Fig.~\ref{fig:km-schematic}.
This is a classic approach to constructing approximate Langevin equations from data~\cite{Friedrich1997prl, Friedrich2011review}.
Boninsegna \textit{et al} also demonstrated that the Kramers-Moyal average can be combined with SINDy sparse regression to discover analytic drift and diffusion equations from data~\cite{Boninsegna2018jcp}.

\subsection{Finite-time effects}
\label{sec:theory-adjoint}

The Kramers-Moyal average has chiefly been a theoretical tool; its application to time series analysis depends strongly on the assumption of Gaussian white noise and fast enough sampling rates to approximate $\tau \rightarrow 0$.
In practice, these two requirements tend to be in tension; if the forcing originates with unresolved scales, then increasing the sampling rate leads to stronger correlations in the ``noise".
The assumption of uncorrelated forcing is never strictly satisfied for continuous physical systems.
Einstein recognized this in his work on Brownian motion~\cite{Einstein1905brownian}, although in that case the separation of scales is pronounced enough that experimental sampling rates run little risk of capturing correlation effects in the molecular forcing.
In complex systems of modern interest, the scale separation is typically much less obvious.

As a simple illustrative model, consider a system driven by a colored noise process $\eta(t)$:
\begin{subequations}
	\label{eq:colored}
	\begin{align}
	\label{eq:colored1}
	\dot{x} &= f(x) + \sigma_x \eta \\
	\label{eq:colored2}
	\dot{\eta} &= -\alpha \eta + \sigma_\eta w(t),
	\end{align}
\end{subequations}
where $w(t)$ is a true Gaussian white noise process.
In this case, the forcing $\eta(t)$ is characterized by a decorrelation time $\alpha^{-1}$.
When $f(x)$ is a stable Navier-Stokes operator linearized about a turbulent mean profile, colored noise forcing has been shown to accurately reproduce turbulent statistics~\cite{Zare2017jfm}.
A generalized Langevin equation driven by time-correlated forcing can also be derived from the Euler equations using the direct-interaction approximation~\cite{Kraichnan1989}.
This perspective is also popular in climate modeling, where models of this form can be derived from the governing equations using  perturbation arguments~\cite{Majda2001cpam}.

Despite the formally non-Markovian nature of the resolved variables, it is advantageous to approximate the forcing as white-in-time, since most foundational theoretical tools are based on this assumption.
One strategy to avoid dealing with the latent variable $\eta(t)$ is deliberate subsampling, as illustrated in Fig.~\ref{fig:scale-separation}.
If the macroscopic dynamics have a characteristic timescale $\omega$, we may be able to choose a sampling rate $f_s = \tau^{-1}$ such that $\omega \ll f_s \ll \alpha$ and the forcing appears decorrelated.
That is, $\langle \eta(t + k\tau) \eta(t) \rangle_t \approx \delta_{k0}$.
In this case the forcing appears to be (band-limited) white noise.

The minimum $\tau$ for which this is true is often called the Einstein-Markov scale~\cite{Friedrich2011review}.
This time scale is not necessarily related to the autocorrelation time of the macroscopic dynamics $x$, but may be identified from data by checking the Markov property at different sampling rates (see App.~\ref{app:sampling-rate} and Ref.~\cite{Friedrich2011review}).
This strategy of subsampling at the approximate Einstein-Markov scale was proposed to model spatial fluctuations in turbulence~\cite{Friedrich1998epl} and the cosmic microwave background~\cite{Ghasemi2006}.

In order to sample the resolved dynamics coarsely enough that the forcing is decorrelated, finite-time effects in the Kramers-Moyal estimates of drift and diffusion must be accounted for~\cite{Ragwitz2001prl}.
However, these effects can be determined exactly in terms of the adjoint Fokker-Planck operator~\cite{Lade2009}.
For notational clarity, we give the result for scalar $x$, although the result generalizes naturally to higher dimensions.

\begin{SCfigure}
	\centering
	\vspace{-.1in}
	\begin{overpic}[width=0.5\linewidth]{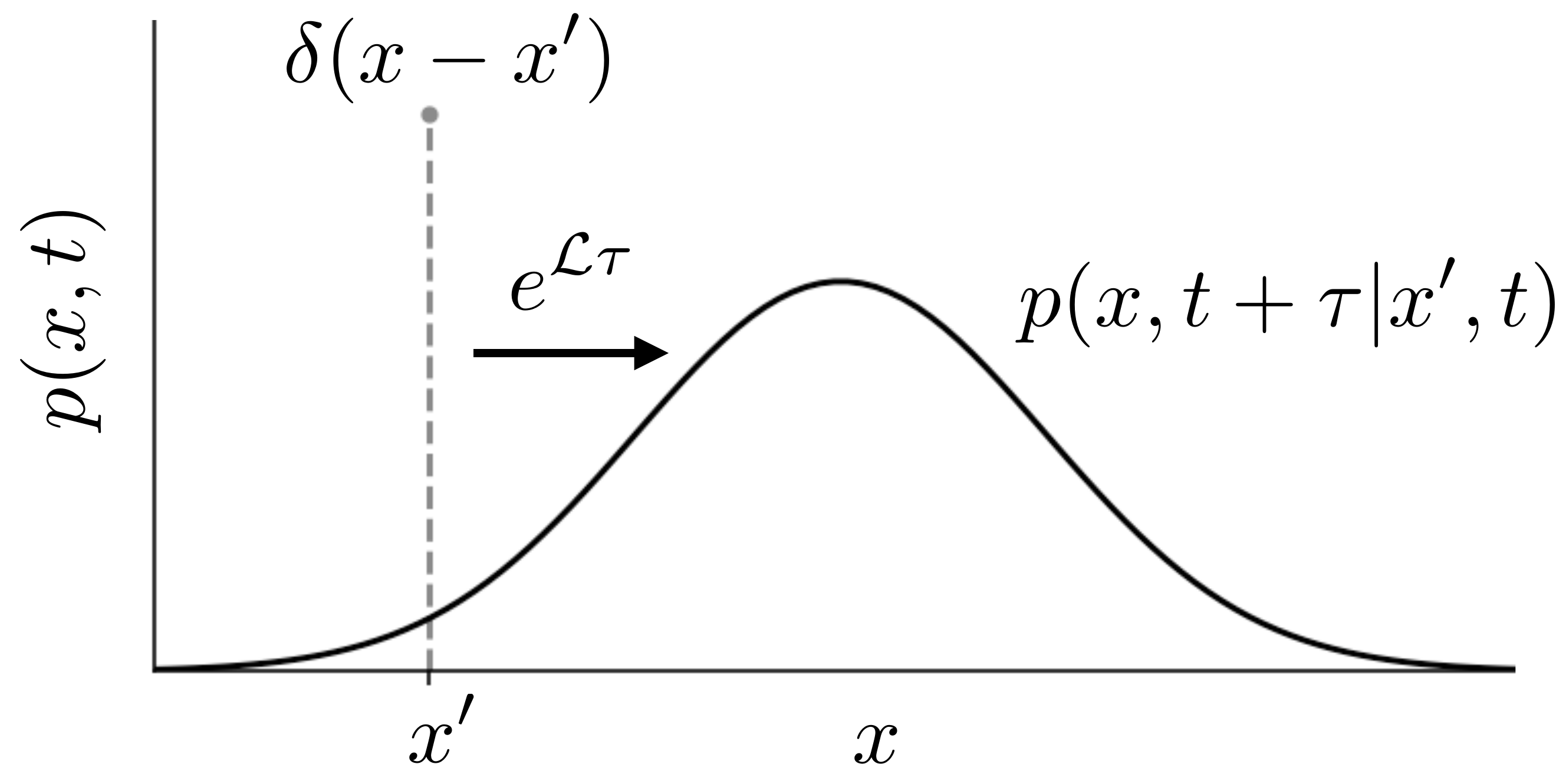}
	\end{overpic}
	\caption{ PDF evolution from the Fokker-Planck operator $\mathcal{L}$.
		The distribution used to evaluate the conditional finite-time moments~\eqref{eq:km-avg-definition} can be interpreted as the evolution of a delta function initial condition over the sampling interval $\tau$, where the state is known to be $x'$ at time $t$.
		\vspace{.2in}
	}
	\label{fig:fp-schematic}
\end{SCfigure}

The conditional moments defined in Eq.~\eqref{eq:km-avg-definition} can be written equivalently as
\begin{equation}
\label{eq:km-int-definition}
m_\tau^{(n)}(x) = \int_{-\infty}^{\infty} (x' - x)^n p(x', t+\tau | x, t) \dd x',
\end{equation}
where $p(x', t+\tau | x, t)$ indicates the conditional joint probability that the system is in state $x'$ at time $t + \tau$ given that it was in state $x$ at time $t$.
If the drift and diffusion are not time-dependent, then the conditional probability $p(x', t+\tau | x, t)$ can be interpreted as the propagation of uncertainty over an interval $\tau$ if the state $x$ is known at time $t$, as shown in Fig.~\ref{fig:fp-schematic}.
According to Eq.~\eqref{eq:forward-fp}, this evolution is given by the Fokker-Planck equation acting on a Dirac delta function:
\begin{equation}
p(x', t+\tau | x, t) = e^{\tau \mathcal{L} (x')} \delta(x'-x).
\end{equation}
Using the definition of the adjoint operator and evaluating the integral with the delta function,
\begin{equation}
\label{eq:adjoint-exact}
m_\tau^{(n)}(x) = \left[ e^{\tau \mathcal{L}^\dagger(x')} (x' - x)^n \right]_{x'=x},
\end{equation}
where the adjoint Fokker-Planck operator is given in tensor summation notation by
\begin{equation} \label{eq:adjoint-fp}
\mathcal{L}^\dagger(x) = f_i(x) \pdv{x_i} + a_{ij}(x) \frac{\partial^2}{\partial x_i \partial x_j}.
\end{equation}

Thus, Lade~\cite{Lade2009} showed that the effect of coarse sampling rates can be understood as the adjoint Fokker-Planck operator evolving the moments of the distribution in time.
Honisch and Friedrich demonstrated the use of this relationship to optimize free parameters in a Langevin model~\cite{Honisch2011pre}; the proposed method in Sec.~\ref{sec:method} builds on this result.

The close correspondence between the Fokker-Planck and Liouville operators also suggests an interpretation of this result in terms of Koopman theory~\cite{Mezic2005,Mezic2013arfm}.
The adjoint Fokker-Planck operator is a \textit{linear} generator of the time evolution of observable functions, even when the underlying dynamics are nonlinear~\cite{Zwanzig2001, Klus2020physica}.
In this case, the observables are the conditional moments $(x'-x)^n$.

These corrections are difficult to compute analytically for nonlinear dynamics, so Eq.~\eqref{eq:adjoint-exact} is typically approximated on a discretized domain.
The matrix exponential $e^{\tau \mathcal{L}^\dagger}$ is relatively inexpensive in one dimension, but more generally it may be helpful to interpret the correction as the solution of a PDE.
That is, if $w^{(n)}(x, t)$ is the solution to the adjoint Fokker-Planck equation
\begin{equation}
\pdv{w^{(n)}}{t} = \mathcal{L}^\dagger(x) w^{(n)}(x, t), \hspace{1cm} w^{(n)}(x, 0) = x^n,
\end{equation}
then by linearity of $\mathcal{L}^\dagger$ the finite-time conditional moments $m_\tau^{(1)}(x)$, $m_\tau^{(2)}(x)$ are given by
\begin{subequations}
	\begin{align}
	m_\tau^{(1)}(x) &= w^{(1)}(x, \tau) - x \\
	m_\tau^{(2)}(x) &= w^{(2)}(x, \tau) - 2 x w^{(1)}(x, \tau) - x^2.
	\end{align}
\end{subequations}


\section{Proposed Langevin regression method}
\label{sec:method}
The decomposition of a multiscale system into dominant deterministic dynamics and stochastic forcing is conceptually simple.  
However, the gulf between detailed first-principles descriptions and a simple Langevin model is large enough that developing accurate stochastic models is difficult.
This is especially true when the dynamics are nonlinear and the unresolved scales have internal dynamics and cannot be treated as true Gaussian white noise.

Due to the intrinsic volatility of individual trajectories, model identification methods typically rely on ergodicity and exploit the connection to ensemble properties via the Fokker-Planck equation.
For example, a scalar model with additive noise can be determined by fitting to the analytic steady-state probability distribution given by Eq. ~\eqref{eq:1d-ss-fp}.
However, this fitting procedure cannot independently estimate drift and diffusion; an additional quantity, such as mean-square displacement, must also be used~\cite{Rigas2015jfm, Brackston2016jfm}.
This presents a challenge for modeling multiplicative noise, which some studies have suggested is important for capturing interactions between the resolved and unresolved degrees of freedom~\cite{Friedrich1997prl, Majda2001cpam}.
Furthermore, the use of a time average to estimate steady-state statistics destroys temporal information, so that oscillatory dynamics cannot be resolved.

Seeking to address a number of these challenges, several recent studies have proposed methods whereby stochastic models may be derived from the Kramers-Moyal average~\cite{Honisch2011pre, Boujo2016prs, Boninsegna2018jcp}.
Here we synthesize and extend these methods into a single optimization framework for identifying sparse, interpretable Langevin-type models from experimental data by constraining the model to both the stationary PDF and the Kramers-Moyal average.
In particular, we generalize the stochastic SINDy method proposed by Boninsegna et al.~\cite{Boninsegna2018jcp} with finite-time corrections that enable modeling systems whose forcing is not approximately given by Gaussian white noise.

\subsection{Known model structure: Fokker-Planck optimization}
\label{sec:methods-opt}

\begin{figure}
	\centering
	\vspace{.05in}
	\begin{overpic}[width=0.95\linewidth]{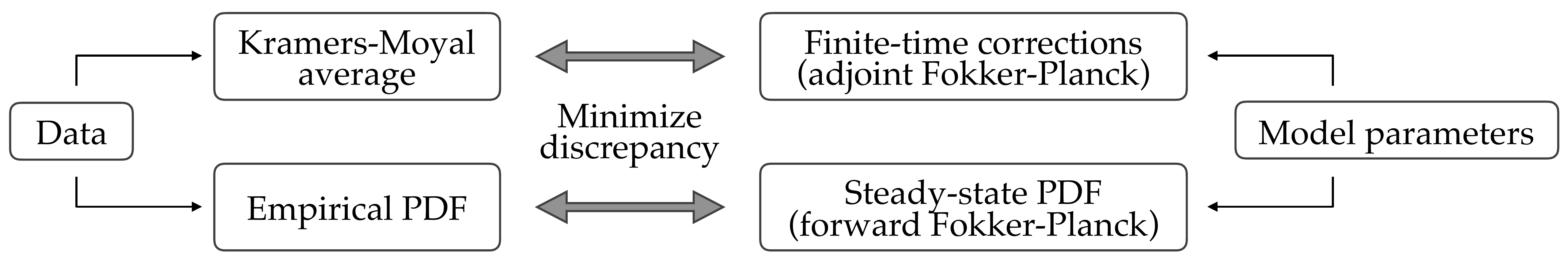}
	\end{overpic}
	\vspace{-.1in}
	\caption{ Schematic of the Langevin regression optimization problem.
		The Fokker-Planck equation can be used to compare the conditional moments and distribution for a proposed model to those observed empirically.
		The model parameters are chosen to minimize the discrepancy between the model and observations, as described in Sec.~\ref{sec:methods-opt}.
		The form of the model may be simultaneously identified with the model selection procedure outlined in Sec.~\ref{sec:methods-sindy}.}
	\label{fig:opt-schematic}
\end{figure}

Estimating the Kramers-Moyal coefficients~\eqref{eq:km-drift} and~\eqref{eq:km-diffusion} by binning a long time series is an attractive option for estimating drift and diffusion functions directly from data.
However, as discussed in Sec.~\ref{sec:theory-adjoint}, the sampling rate must be chosen carefully for unresolved dynamics to decorrelate.
This subsampling can introduce significant finite-time distortion to the empirical Kramers-Moyal coefficients.
This distortion is given in terms of the adjoint Fokker-Planck equation by Eq.~\eqref{eq:adjoint-exact}, although constructing the Fokker-Planck operator itself requires the drift and diffusion.

Based on these considerations, Honisch \& Friedrich suggested an iterative procedure to estimate free parameters of drift and diffusion functionss~\cite{Honisch2011pre}, which may be summarized as follows.
If the Langevin model is given in terms of a set of parameters $\xi$, so that
\begin{equation}
\dot{x} = f(x; \xi) + \sigma(x; \xi) w(t),
\end{equation}
then the problem is to choose the parameters such that the finite-time conditional moments are consistent with the empirical Kramers-Moyal estimates.
Leaving aside numerical details, the optimization consists of the following:
\paragraph{Parameter optimization}
\begin{enumerate}
	\item Select an appropriate sampling rate $\tau$ (see App.~\ref{app:sampling-rate});
	\item Estimate empirical finite-time conditional moments $\hat{m}_\tau^{(n)}(x)$ for $n=1, 2$ using Eq.~\eqref{eq:km-avg-definition};
	\item For a set of parameters $\xi$, construct the adjoint Fokker-Planck operator $\mathcal{L}^\dagger$ in Eq.~\eqref{eq:adjoint-fp};
	\item Compute the exact moments $m_\tau^{(n)}(x, \xi)$ using the adjoint correction given by Eq.~\eqref{eq:adjoint-exact};
	\item Choose $\xi$ to minimize the discrepancy between the empirical moments $\hat{m}_\tau^{(n)}(x)$ and the exact moments $m_\tau^{(n)}(x, \xi)$.
\end{enumerate}
More concretely, the optimal $\xi$ solves the following problem on a discrete domain of $N$ points $x_i$:
\begin{equation}
\label{eq:afp-opt}
\min_{\xi} \sum_{n=1}^2 \sum_{i=1}^N w_i^{(n)} \left[m_\tau^{(n)} (x_i, \xi) - \hat{m}_\tau^{(n)} (x_i) \right]^2.
\end{equation}
Here the weights $w_i^{(n)}$ reflect pointwise uncertainty in the empirical estimate of the moments.

Due to the diffusive nature of the Fokker-Planck equation, it is not clear that this problem is necessarily well-posed.
That is, when the system is sampled coarsely there may be a range of parameters that are consistent with the observed conditional moments within experimental uncertainty.

We propose ``regularizing" the optimization problem~\eqref{eq:afp-opt}, as proposed by Ref.~\cite{Honisch2011pre}, with the Kullbeck-Leibler (KL) divergence $\mathcal{D}_\mathrm{KL}$ between the empirical PDF $\hat{p}(x)$ and the solution of the steady-state Fokker-Planck equation $p(x, \xi)$, given by Eq.~\eqref{eq:forward-fp}.
The modified cost function is
\begin{equation}
\label{eq:opt-reg}
\min_{\xi} \sum_{n=1}^2 \sum_{i=1}^N w_i^{(n)} \left[m_\tau^{(n)} (x_i, \xi) - \hat{m}_\tau^{(n)} (x_i) \right]^2 + \eta \mathcal{D}_\mathrm{KL} \left( \hat{p}(x), p(x, \xi \right) ),
\end{equation}
where $\eta$ is the relative weight of the two contributions to the cost function.
The KL divergence is a statistical measure of the difference between two probability distributions $p$ and $q$, defined as
\begin{equation}
\label{eq:kl-div}
\mathcal{D}_\mathrm{KL} (p, q) = \int p(x) \log \left( \frac{p(x)}{q(x)} \right) \dd x.
\end{equation}
This regularization ensures that the resulting Langevin model is consistent with both the finite-time Kramers-Moyal coefficients and the asymptotic steady-state probability distribution.
The optimization problem in Langevin regression is shown schematically in Fig.~\ref{fig:opt-schematic}.
Because the typical dimension of $\xi$ is relatively small compared to the dimension $x$, we find that gradient-free optimization methods such as a Nelder-Mead simplex search are more efficient than those designed for large parameter spaces and relatively inexpensive cost function evaluations, such as automatic differentiation.

The resulting optimization problem requires solution of both the forward and adjoint Fokker-Planck equations at each evaluation of the cost function.
For a scalar variable $x$, the steady-state solution to the forward equation can be computed directly with Eq.~\eqref{eq:1d-ss-fp}.
In higher dimensions, there is no analytic steady-state solution to the forward equation; we describe several solvers in App.~\ref{app:ss-FP}.
We solve the adjoint equation with a second-order finite difference method.
In one and two dimensions, the resulting matrix exponential is relatively inexpensive, while for higher dimensions a time-stepping approach that exploits the sparse operator structure will be more efficient.

\subsection{Unknown model structure: stochastic SINDy}
\label{sec:methods-sindy}

If the form of the model can be assumed up to a set of unknown parameters, the Fokker-Planck optimization problem detailed in the previous section is sufficient to estimate the free parameters.
However, in many cases we might have some partial prior assumptions about the model structure (e.g. the model consists of polynomials with a particular symmetry, or that one variable is forced by another), but the exact form is unknown.
The sparse identification for nonlinear dynamics (SINDy) method has recently shown promise for obtaining nonlinear reduced-order models  of laminar flows~\cite{Brunton2016pnas, Loiseau2017jfm, Loiseau2019, Deng2020jfm} from data.  
However, SINDy typically relies on estimated time derivatives, which is a significant barrier to modeling experimental data or multiscale systems. 
In related work, SINDy has recently been leveraged for turbulence closure modeling~
\cite{beetham2020formulating}.   

Boninsegna \textit{et al}~\cite{Boninsegna2018jcp} recently proposed a stochastic SINDy algorithm based on the Kramer-Moyal average without an adjoint correction for finite-time effects.
Empirical estimation of conditional moments gives point estimates of drift and diffusion at each histogram bin; if the moments are estimated reliably, then the system identification problem reduces to fitting a curve through these points.
In its simplest form, a parsimonious model can be chosen using the SINDy framework, where the model parameters are a coefficient vector for a "library" matrix whose columns consist of candidate functions~\cite{Brunton2016pnas}.
For instance, the library $\boldsymbol{\Theta}(x)$ might consist of polynomials in $x$; then we look for polynomial representations of the drift $f(x)$ and diffusion $\sigma(x)$, so that
\begin{equation}
\label{eq:sindy-representation}
f(x) = \boldsymbol{\Theta}^T_f(x) \xi_{f}  \hspace{2cm} \sigma(x) =  \boldsymbol{\Theta}^T_\sigma(x) \xi_{\sigma},
\end{equation}
where $\xi_f$ and $\xi_\sigma$ are sparse vectors that have as many zero entries as possible while still capturing the observed dynamics.
Standard sparse regression algorithms can be used to select a set of functions balancing parsimony and accuracy.

In the deterministic SINDy algorithm, this regression problem is constructed by concatenating column vectors of an estimated time derivative $\dot{x}$ and the evaluations of the candidate functions.
In the present case, however, the regression is performed over the discretized spatial domain rather than a long time series.
The conditional finite-time coefficients are estimated by computing Eq.~\eqref{eq:km-avg-definition} over observations that fall into each spatial histogram bin, as visualized in Fig.~\ref{fig:km-schematic}.
The regression problem is constructed over these bins rather than the direct time series.
In practice this typically reduces the length of the column vectors from $\mathcal{O}(10^6)$ to $\mathcal{O}(10^2)$.

One consequence of this formulation is that standard sparse regression algorithms, such as thresholded least squares or forward regression orthogonal least squares, do not work well.
A simple alternative is the reverse-greedy stepwise sparse regression (SSR)~\cite{Boninsegna2018jcp}.
With this method, terms are sequentially removed from the model according to some criteria.
In the original SSR algorithm, the coefficients were identified with a simple least squares and terms with the smallest absolute value were removed.
However, in general the smallest coefficient does not necessarily imply the least important contribution.
For this reason we sequentially remove terms corresponding to the smallest \textit{increase} in cost function.
The cost function itself can then serve as a model-selection criterion; for a Pareto-optimal model the cost function should jump significantly from a near-minimum value once important terms begin to be discarded.

When combined with the forward/adjoint Fokker-Planck optimization described in the previous section, this model selection procedure represents a flexible and general framework for identifying stochastic approximations to multiscale nonlinear dynamics, which we refer to as \textit{Langevin regression}.
In the following section, the Fokker-Planck parameter estimation is demonstrated on two example systems for which the form of the model is clear.
The ability of Langevin regression to simultaneously identify the structure and parameters of a model from data is demonstrated in Sec.~\ref{sec:results-wake} for experimental measurements of a turbulent wake.


\section{Results}
\label{sec:results}

Here we apply Langevin regression to three example problems of increasing complexity.
First, we show that the coarse sampling and scale separation ideas discussed in Sec.~\ref{sec:theory} enable the identification of stochastic systems with time-correlated forcing.
For this example, we assume knowledge about the structure of the model in order to highlight the effects of colored noise and parameter estimation in the case where the correct structure is known.

Second, we demonstrate the construction of a statistically consistent reduced-order model by approximating the second-order dynamics of a particle in a double-well potential with the corresponding first-order bifurcation normal form.
A stochastic normal form model can be derived analytically for this system, although its accuracy quickly degrades away from the bifurcation point.
We fix the structure of the model and show that Langevin regression can maintain statistical accuracy even far from the bifurcation point.

Finally, we derive an accurate and efficient stochastic model for the turbulent flow in the wake of an axisymmetric bluff body from experimental measurements.  
This example presents several challenges, including partial and noisy measurements of a multiscale system, and has relevance to numerous industrial applications~\cite{Brunton2015amr}.  
Although the structure of the drift dynamics may be inferred from laminar stability analysis, we instead apply sparse model selection to discover this structure entirely from data.
Our procedure identifies a simple and interpretable nonlinear model with a multiplicative noise term that improves the correspondence with the empirical power spectrum and probability distribution compared with previous stochastic modeling results.

The synthetic examples in Sec.~\ref{sec:results-pitchfork} and~\ref{sec:results-doublewell} are simulated using the SRIW1 stochastic Runge-Kutta method~\cite{Rossler2010sde}, available in the DifferentialEquations.jl package~\cite{DifferentialEquations.jl}.
Langevin regression is performed on Kramers-Moyal coefficients and empirical PDFs computed from a time series of $10^7$ points sampled at $\Delta t = 10^{-2}$.
The turbulent wake model in Sec.~\ref{sec:results-wake} is based on the aerodynamic center of pressure, a global integral quantity estimated from 64 evenly spaced pressure taps~\cite{Rigas2014jfm}.
The model is estimated from a time series of $8.9 \times 10^6$ experimental measurements of the center of pressure sampled at 225 Hz.
Monte Carlo evaluation of this model is performed in Python using a standard Euler-Maruyama numerical integration scheme at the same sampling rate.
The coarse subsampling rate for Kramers-Moyal averaging is chosen for each system according to the criteria discussed in App.~\ref{app:sampling-rate}.
The relative weight $\eta$ of the Kullback-Leibler divergence in the optimization function is a multiple of 10, which is chosen to be roughly equal to the minimum cost function with $\eta = 0$.
In other words, the Kramers-Moyal coefficients and the PDF are given roughly equal weight in the optimization.

\subsection{Pitchfork bifurcation normal form}
\label{sec:results-pitchfork}
\vspace{-0.05in}

\begin{figure}
	\centering
	\begin{overpic}[width=0.85\linewidth]{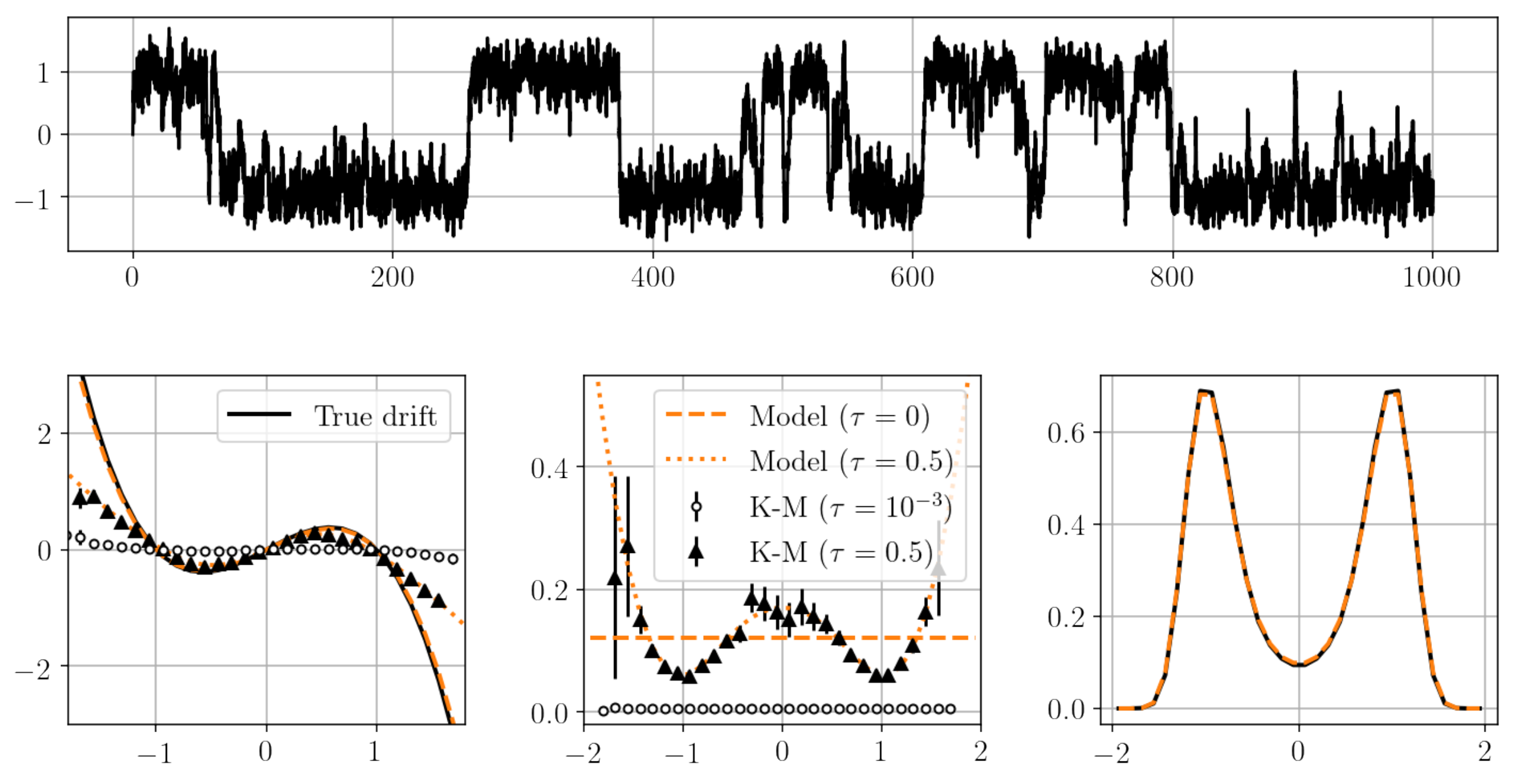}
		\small{
			\put(-2, 42){\begin{sideways}
					\large{$x$} \end{sideways}}
			\put(52, 30){\large{$t$}}
			\put(-2, 12){\begin{sideways} \large{$f(x)$} \end{sideways}}
			\put(31, 12){\begin{sideways} \large{$a(x)$} \end{sideways}}
			\put(66, 12){\begin{sideways} PDF \end{sideways}}
			\put(16.5, -1.5){\large{$x$}}
			\put(50.5, -1.5){\large{$x$}}
			\put(84.5, -1.5){\large{$x$}}
		}
	\end{overpic}
	\caption{ 
		\small{\textbf{Pitchfork normal form forced by colored noise.}
			Top: the noise induces random switching between the metastable equilibria.
			Bottom: if the Kramers-Moyal coefficients are computed with a sampling rate faster than the decorrelation of the noise ($\circ$), the drift still appears cubic but the amplitude is underestimated by approximately an order of magnitude. On the other hand, if the noise is allowed to decorrelate ($\triangle$), the estimated Kramers-Moyal coefficients are the right order of magnitude, but are distorted from the zero-time value. The diffusion appears multiplicative and quadratic.
			The adjoint finite-time corrections recover a consistent Langevin model driven by white noise (\textcolor{mpl_orange}{$- -$}).
		}
	}
	\label{fig:pitchfork}
\end{figure}
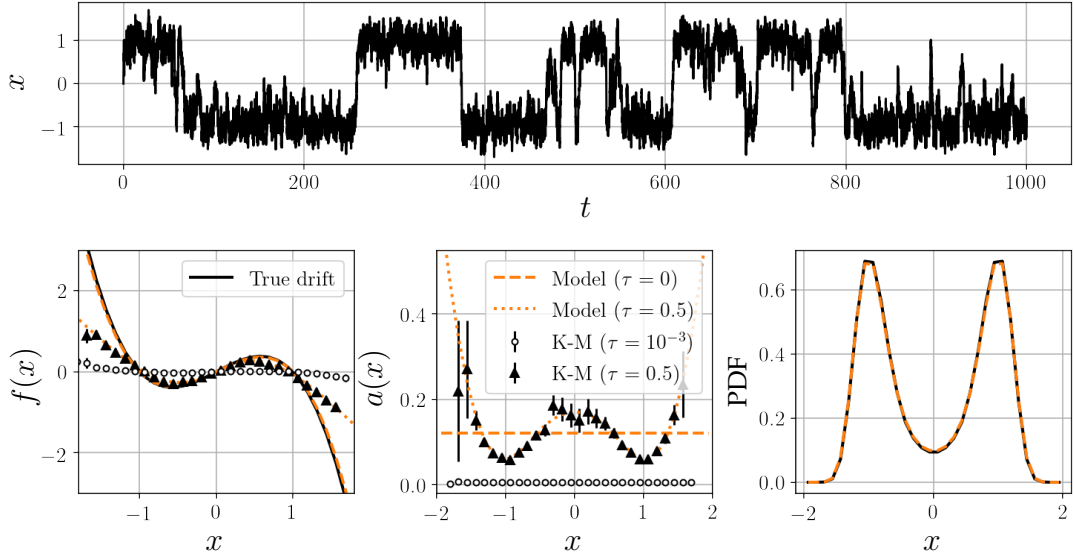

The normal form for a supercritical pitchfork bifurcation provides a canonical example of bistability, where an eigenvalue with zero imaginary part crosses the real axis as a parameter is varied.
For example, this normal form describes the amplitude equation governing the symmetry-breaking mode of the wake behind a circular disk, which can be derived by a weakly nonlinear stability analysis~\cite{Meliga2009jfm}.
This result inspired the use of a stochastically forced pitchfork normal form to model the turbulent evolution of the centroid of the base pressure distribution on the back of an axisymmetric bluff body~\cite{Rigas2014jfm} and the bistability of a three-dimensional Ahmed body wake~\cite{Brackston2016jfm}.

However, unresolved degrees of freedom in a turbulent flow do not typically resemble delta-correlated white noise. 
Here we investigate the impact of correlated noise on stochastic system identification by considering the supercritical pitchfork normal form forced by colored noise:
\begin{subequations}
	\begin{align}
	\label{eq:pitchfork-nf}
	\dot{x} &= \lambda x - \mu x^3 + \eta \\
	\label{eq:ornstein-uhlenbeck}
	\dot{\eta} &= -\alpha \eta + \sigma w(t).
	\end{align}
\end{subequations}
Here $\eta$ is an Ornstein-Uhlenbeck process with characteristic relaxation time $\alpha^{-1}$, which acts as an effective low-pass filter on the white noise process $w(t)$.
We choose $\mu = \lambda = \beta = 1,$ $\alpha = 10^2$, and $\sigma = 0.5 \alpha $ so that the typical amplitude of $\eta$ is around $0.5$.

The nonzero relaxation rate for the Ornstein-Uhlenbeck process introduces temporal correlations that invalidate many of the standard analytic approaches to stochastic modeling if $\eta$ is not directly observed.
However, if the relaxation rate is much larger than any natural dynamics of the slow variable (i.e. $\alpha \gg \lambda$) we can appeal to the dual scale separation idea of Fig.~\ref{fig:scale-separation} and approximate the fast scales as uncorrelated noise.

As demonstrated in Fig.~\ref{fig:pitchfork}, the correlated forcing destroys the Kramers-Moyal average as the sampling interval $\tau \rightarrow 0$.
This can be mitigated by sampling coarsely enough that the noise decorrelates and appears to whiten.
For instance, if we choose $\tau = 0.5 = 50 \alpha = 0.5\lambda$ (slower than the noise decorrelation but faster than the drift dynamics), the Kramers-Moyal average is of the correct order of magnitude.
However, finite-time sampling rates now significantly deform the observed drift and diffusion, even introducing apparent state dependence in the diffusion (Fig.~\ref{fig:pitchfork}, bottom middle).
This observation by Ragwitz \& Kantz called into question many earlier attempts to use the Kramers-Moyal average for modeling without accounting for finite-time effects~\cite{Ragwitz2001prl}.

Langevin regression accounts for these distortions with the adjoint Fokker-Planck operator, recovering a one-dimensional model nearly identical to Eq.~\eqref{eq:pitchfork-nf}, but forced by additive white noise.
The identified coefficients (given in Table~\ref{tab:pitchfork}) differ from the true values by around 5\%, but the model closely matches both the observed finite-time conditional moments and the empirical probability distribution (Fig.~\ref{fig:pitchfork}, bottom row).
This suggests that the proposed subsample-and-correct approach is capable of identifying statistically consistent Langevin models, even in the presence of correlated noise.
This result depends fundamentally on the dual scale separation principle of Fig. \ref{fig:scale-separation}; the success of this approach may be limited when these timescales cannot be clearly separated with the coarse sampling rate.

\begin{table}[h!]
	\centering
	\begin{tabular}{||c c c c||} 
		\hline
		Model & $\lambda$ & $\mu$ & $\sigma$ \\
		\hline\hline
		True system (colored noise) & 1.0 & 1.0 & $--$ \\
		Stochastic SINDy (no adjoint) & 0.43 & 0.43 & 0.44 \\
		Langevin regression (white noise) & 0.96 & 0.96 & 0.49 \\
		\hline
	\end{tabular}
	\caption{\small{True parameters for the pitchfork normal form forced by colored noise along with those estimated from data. Without the adjoint Fokker-Planck corrections for finite sampling rates, the true coefficients are underestimated by a factor of 2. On the other hand, Langevin regression with full adjoint-based optimization identifies a statistically consistent model driven by white noise forcing, where the drift coefficients are a close match to the true system.}}
	\label{tab:pitchfork}
\end{table}

\subsection{Double-well potential}
\label{sec:results-doublewell}
\vspace{-0.05in}

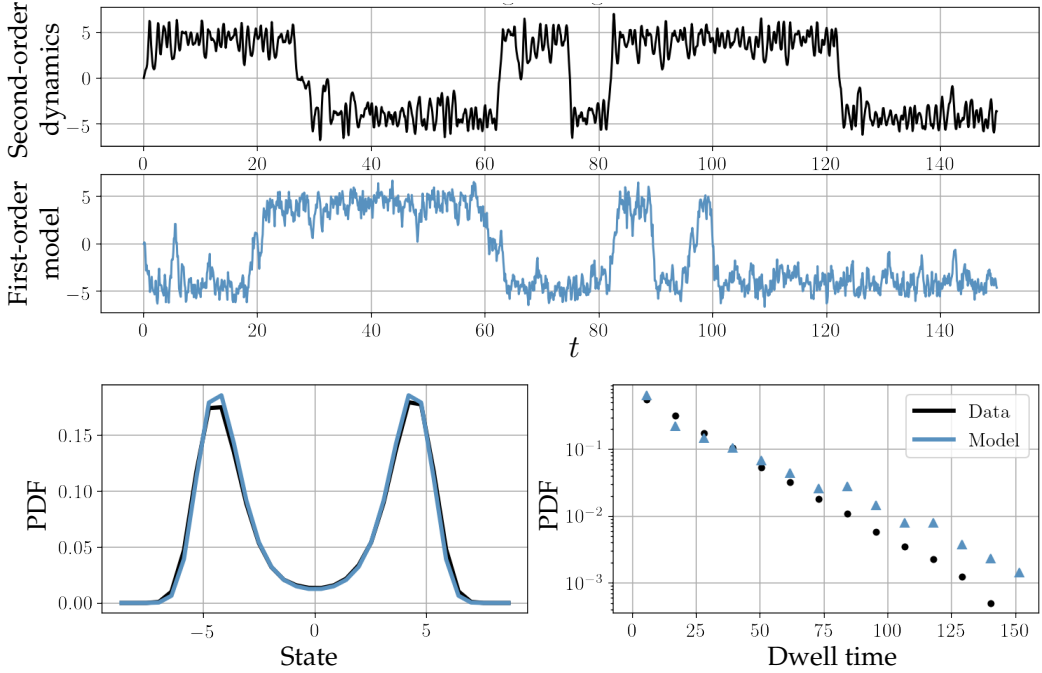
\begin{figure}
	\centering
	\begin{overpic}[width=0.8\linewidth]{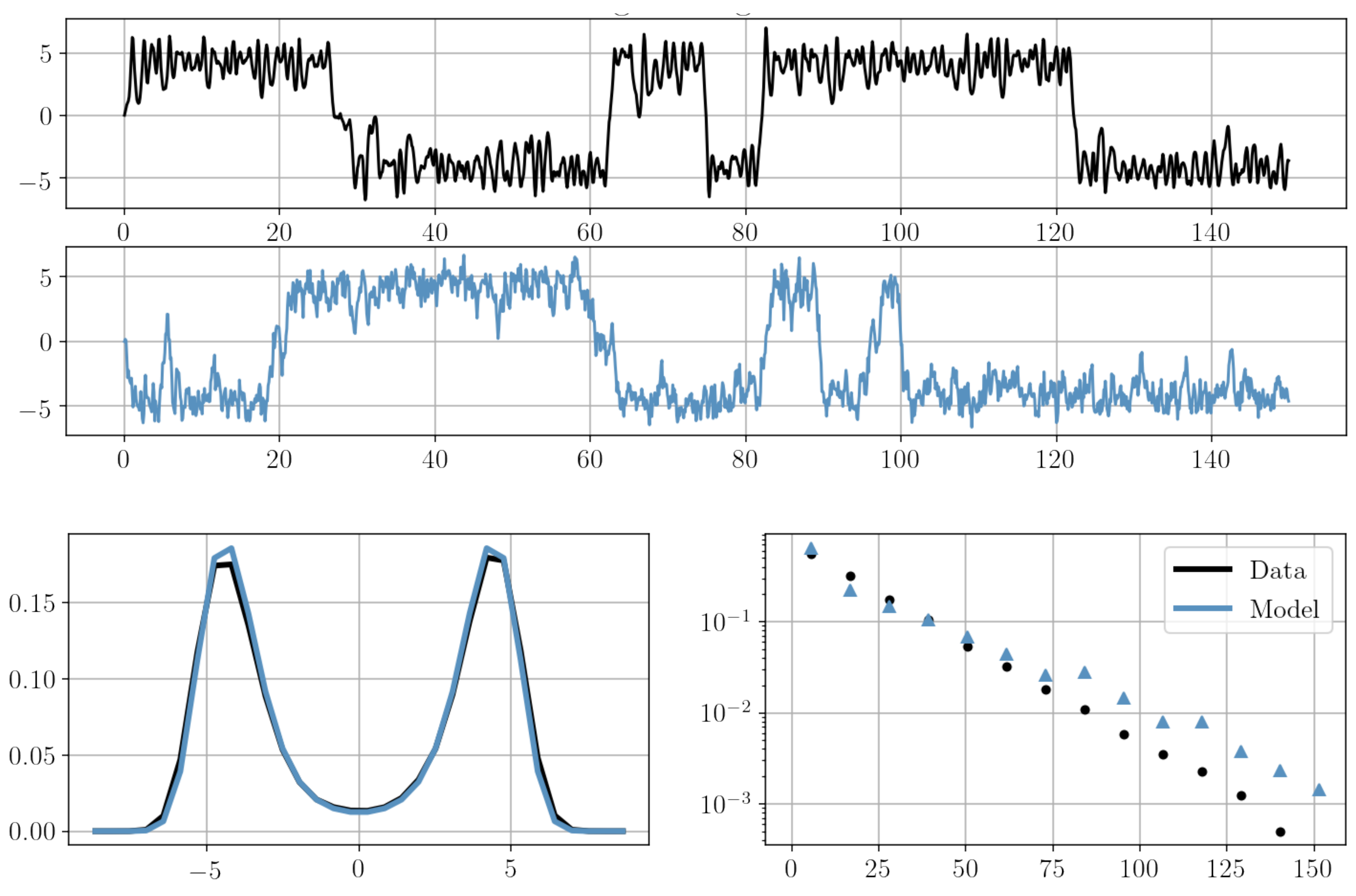}
		\small{
			\put(-4.5, 49.5){\begin{sideways} \parbox{60pt}{\centering Second-order \\ dynamics }\end{sideways}}
			\put(-4.5, 33){\begin{sideways} \parbox{60pt}{\centering First-order \\ model }\end{sideways}}
			\put(-2.5, 12){\begin{sideways} PDF \end{sideways}}
			\put(52, 29.5){\large{$t$}}
			\put(23, -1.5){State}
			\put(49, 12){\begin{sideways} PDF \end{sideways}}
			\put(72, -1.5){Dwell time}
		}
	\end{overpic}
	\caption{ 
		\small{\textbf{Particle in a one-dimensional double-well potential.}
			Even when the system is far from the pitchfork bifurcation ($\epsilon = 20$), the dynamics are dominated by bistable switching behavior (top).
			Langevin regression identified a reduced-order approximation to this system (middle), which is consistent with both the state probability distribution (bottom left) and the distribution of metastable dwell times (bottom right).
		}
	}
	\label{fig:doublewell}
\end{figure}


In many cases, Langevin-type stochastic models are intended to be reduced-order approximations of the large-scale dynamics of a complex system, rather than faithful representations of first-principles physics.
The ``microscopic" degrees of freedom in these systems generally have finite correlation times, as with the colored noise in the previous example.
Eliminating these variables from the model leads to explicit memory effects in the Langevin equations~\cite{Zwanzig2001}, unless the scale separation principle can be employed to identify a  memory-free reduced-order stochastic model from data.
Low-order polynomial dynamics, such as normal forms, can arise naturally in this context as a way to describe the macroscopic behavior, even when the underlying physical description appears completely different~\cite{GuckenheimerHolmes}.

For example, the one-dimensional motion of a particle of unit mass in a general potential $U(x)$ subject to thermal fluctuations is given by
\begin{equation}
\ddot{x} + \gamma \dot{x} + U'(x) = \sqrt{2 \gamma \kB T} w(t),
\end{equation}
where $\gamma$ is the damping ratio, $\kB$ is Boltzmann's constant, $T$ is the temperature, and $w(t)$ is a white noise process~\cite{Risken1996book}.
We consider the double-well potential
\begin{equation}
U(x) = -\frac{\alpha}{2} x^2 + \frac{\beta}{4} x^4.
\end{equation}
An example trajectory of this system is shown in Fig.~\ref{fig:doublewell}, displaying both small oscillations within each well and random large jumps between wells.

The equation of motion generated by this potential can be nondimensionalized to the form
\begin{equation}
\label{eq:doublewell-nondim}
\dv{t} \begin{bmatrix} x \\ \dot{x} \end{bmatrix}
= \begin{bmatrix} 0 & 1 \\
\epsilon - x^2 & -2 \end{bmatrix}
\begin{bmatrix} x \\ \dot{x} \end{bmatrix} + \begin{bmatrix} 0 \\ \sigma \end{bmatrix} w.
\end{equation}
The drift undergoes a supercritical pitchfork bifurcation at $\epsilon = 0$, where the origin loses stability to the pair of fixed points $ x = \pm \sqrt{\epsilon}$.
For small $\epsilon$, an invariant manifold approximation (see App.~\ref{app:normal-form}) gives a first-order model based on the pitchfork bifurcation normal form:
\begin{equation}
\label{eq:doublewell-nf}
\dot{x} = \lambda(\epsilon) x - \mu(\epsilon) x^3 + \tilde{\sigma} w(t).
\end{equation}

As Fig.~\ref{fig:doublewell} shows for $\epsilon = 20$, even far from the bifurcation the bistability still dominates the dynamics, although the invariant manifold approximation used to reduce the order of the normal form model no longer holds.
A first-order equation of the form of Eq.~\eqref{eq:doublewell-nf} can capture this bistability at the cost of ignoring the small oscillations within each potential well.
In other words, the goal is to coarse-grain the dynamics while preserving the statistical properties of the system.

\begin{SCfigure}
	\centering
	\begin{overpic}[width=0.6\linewidth]{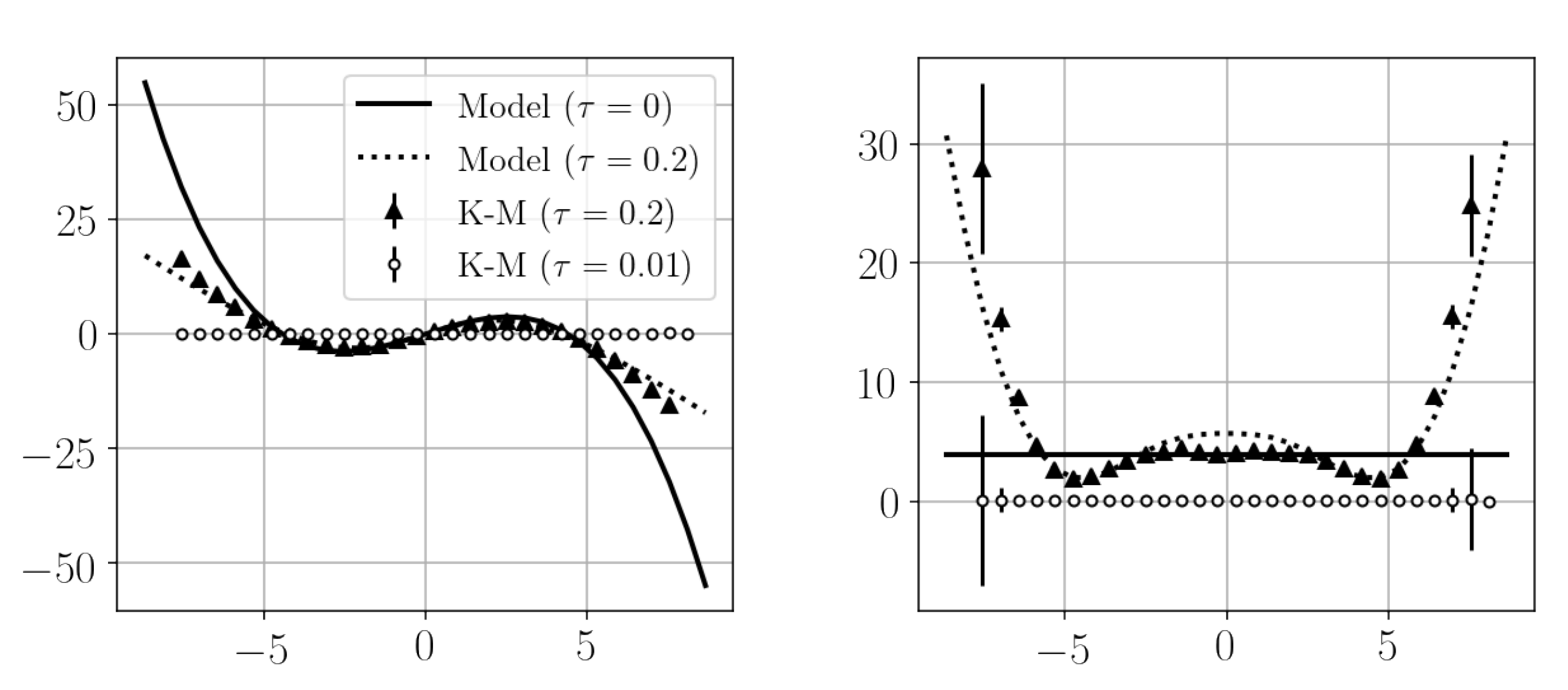}
		\small{
			\put(-3, 15){\begin{sideways} Drift $\hat{f}_\tau (x)$ \end{sideways} }
			\put(49.5, 12){\begin{sideways} Diffusion $\hat{a}_\tau (x)$ \end{sideways} }
			\put(25.8, -1){$x$}
			\put(77.2, -1){$x$}
		}
	\end{overpic}
	\caption{ 
		\small{\textbf{Finite-time effects in Kramers-Moyal average.}
			Non-Markovian effects from order reduction cause the Kramers-Moyal average to fail at high sampling rates ($\circ$).
			The forcing appears uncorrelated when the system is subsampled ($\triangle$). 
			Langevin regression identifies a model (\textbf{---}) which is consistent with the Kramers-Moyal average at finite sampling rate (\textbf{- -}).
		}
	}
	\label{fig:doublewell-km}
\end{SCfigure}
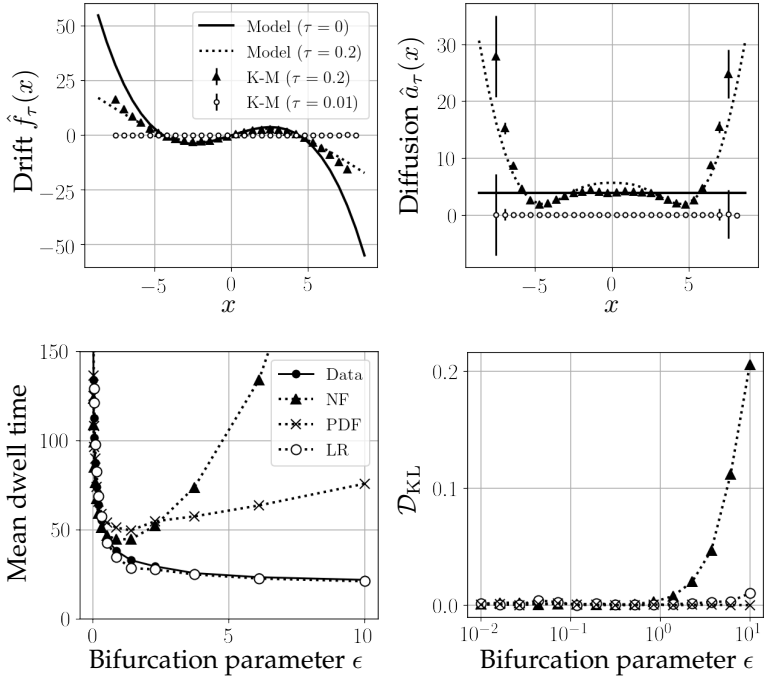

Constructing a first-order Langevin model for the position $x$ is made difficult by the fact that the time series is smoothed by integration of the thermal fluctuations forcing $\dot{x}$ in Eq.~\eqref{eq:doublewell-nondim}.
The neglected degree of freedom introduces non-Markovian behavior and confounds the Kramers-Moyal average, which tends towards zero with fast sampling rates (Fig.~\ref{fig:doublewell-km}).
However, by choosing a coarse enough sampling rate so the subsampled dynamics appear Markovian and correcting for the finite-time effects, Langevin regression is able to identify a first-order model that captures both the probability distribution $p(x)$ and the distribution of residence times in each metastable well.

\begin{SCfigure}
	\centering
	\begin{overpic}[width=0.6\linewidth]{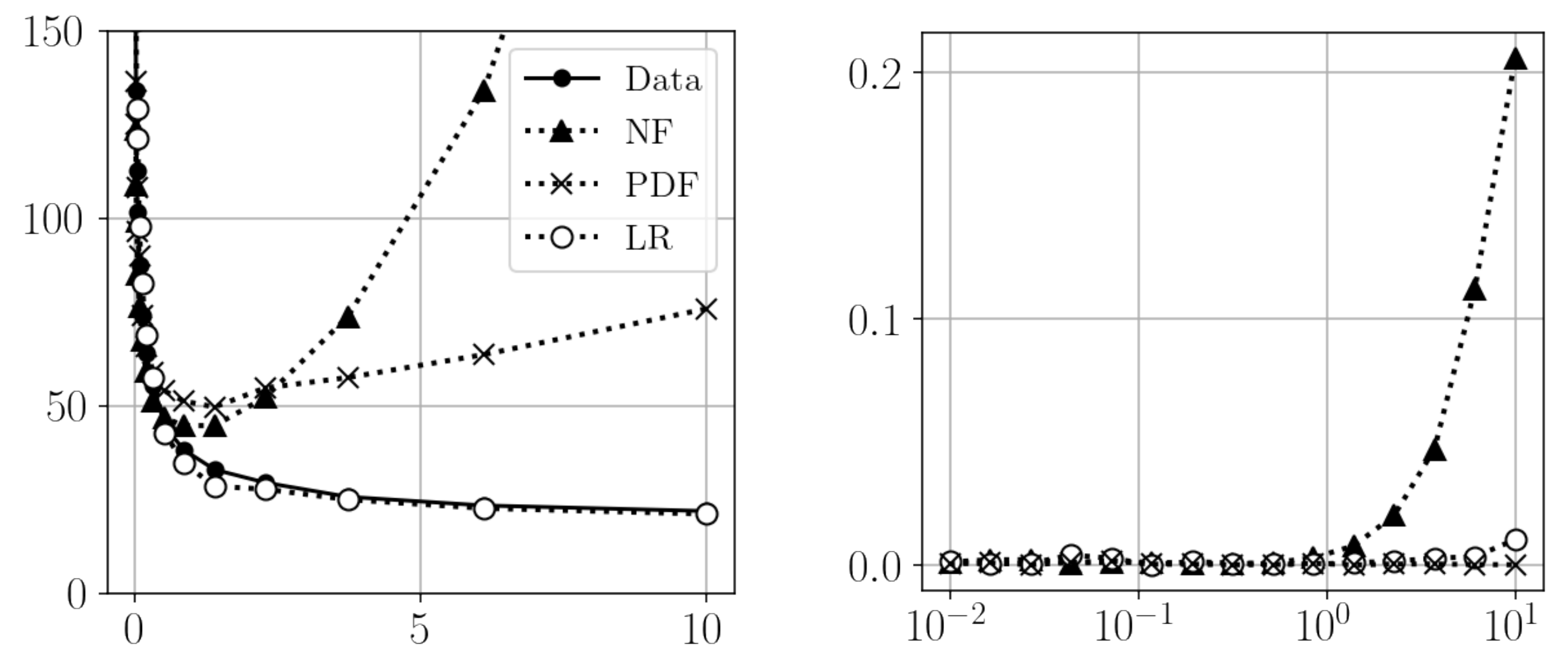}
		\small{
			\put(-3, 9){\begin{sideways} Mean dwell time \end{sideways} }
			\put(49.5, 18){\begin{sideways} $\mathcal{D}_\mathrm{KL}$ \end{sideways} }
			\put(8, -2.5){Bifurcation parameter $\epsilon$}
			\put(60.5,-2.5){Bifurcation parameter $\epsilon$}
		}
	\end{overpic}
	\vspace{.2in}
	\caption{ 
		\small{\textbf{Comparison of models far from bifurcation with Monte Carlo simulation.}
			Both the normal form (NF) and the PDF fit described in App.~\ref{app:ss-FP-1d} quickly fail to match the metastable dwell time, while the Langevin regression (LR) model continues to be accurate far from the bifurcation.
			The Kullback-Leibler divergence $\mathcal{D}_\mathrm{KL}$ measures the difference between the model and system probability distribution.
		}
	}
	\label{fig:doublewell-nf}
\end{SCfigure}

As shown in Fig.~\ref{fig:doublewell-nf}, the Langevin regression model has similar fidelity to the analytic normal form close to the bifurcation, but the data-driven model maintains statistical accuracy well beyond the region where the normal form is valid.
Also shown are results for the same model structure, but with parameters estimated by regression to the empirical PDF (App.~\ref{app:ss-FP-1d}).
At each value of the bifurcation parameter $\epsilon$, the second-order dynamics are driven by noise $\sigma = (\sqrt{\epsilon} + \epsilon)/2$ so that the mean dwell time maintains a similar order of magnitude throughout the range of comparison.

\subsection{Turbulent axisymmetric wake}
\label{sec:results-wake}

Turbulence is a notoriously challenging problem that exemplifies many of the difficulties of stochastic modeling. A turbulent fluid flow is deterministic in principle, but the large and continuous range of spatiotemporal scales often necessitates statistical analysis.
However, unlike Brownian motion, turbulence is far enough from equilibrium that it does not obey the principle of detailed balance; the machinery of statistical mechanics cannot easily be applied to turbulence~\cite{Kraichnan1989}.
Nevertheless, many turbulent flows are dominated by large-scale coherent structures whose regular evolution is suggestive of low-dimensional dynamics, despite the unpredictability introduced by strong coupling to the smaller scales in the flow.
In particular, high Reynolds number flows are characterized by a wide separation between the slow macroscopic dynamics and the faster turbulent fluctuations~\cite{Tennekes1972book}.

\begin{figure}
	\centering
	\vspace{.3in}
	\begin{overpic}[width=0.9\linewidth]{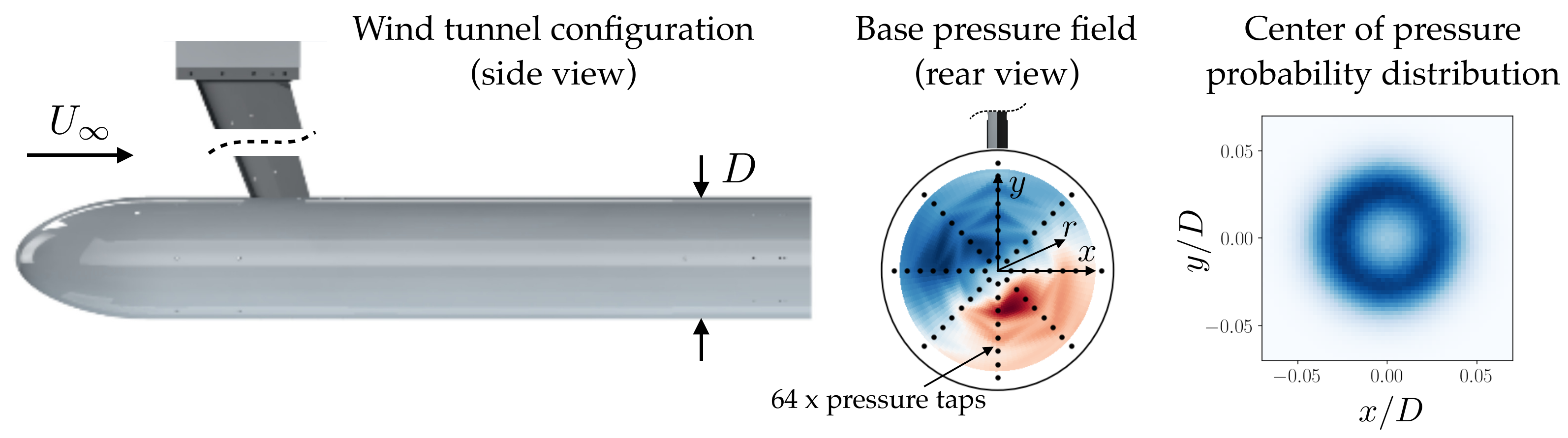}
	\end{overpic}
	\caption{ \small{\textbf{Experimental configuration for the axisymmetric wake.}
			The bluff body is mounted from the wind tunnel ceiling (left) and the base pressure distribution is measured from 64 pressure taps (middle).
			The symmetry-breaking instability of the laminar flow persists in the fully turbulent wake, although the center of pressure appears to wander randomly, as seen in the probability distribution $p(x, y)$ (right).
		}
	}
	\label{fig:wake-config}
\end{figure}

In this context, Langevin regression is a natural extension of the data-driven modeling methods that have proven successful at identifying low-dimensional dynamics in laminar flows~\cite{Loiseau2017jfm,Loiseau2018jfm, Loiseau2019, Deng2020jfm}.
Just as these methods are capable of generalizing the near-bifurcation results of weakly nonlinear analyses~\cite{Sipp2007jfm, Meliga2009jfm} and POD-Galerkin models~\cite{Noack2003jfm}, here we aim to further extend this philosophy to turbulent flows by modeling all but the most important degrees of freedom as stochastic forcing.

We demonstrate stochastic model identification on the experimentally measured pressure distribution on an axisymmetric bluff body, visualized in Fig.~\ref{fig:wake-config} and described in detail in Ref.~\cite{Rigas2014jfm}.
The Reynolds number based on the body diameter is $\Re \sim 2 \times 10^5$.
This flow is a stereotypical configuration that exhibits several features important to drag reduction applications.

The low Reynolds number laminar wake behind an axisymmetric bluff body is symmetric, but undergoes a supercritical pitchfork bifurcation at $\Re \sim 10^2$ so that the center of pressure is offset at a nonzero radial amplitude~\cite{Meliga2009jfm}.
Since the unstable symmetric wake has lower drag than the asymmetric configuration, stabilizing the symmetric state is a major goal of flow control studies~\cite{Beaudoin2006pof,Brackston2016jfm}.
Simple low-dimensional models that accurately represent process noise, energy transfers, and frequency dynamics could significantly improve closed-loop control schemes.
The symmetry-breaking instability continues to dominate the wake dynamics in the turbulent regime, although the location of the center of pressure tends to wander randomly~\cite{Grandemange2013jfm, Rigas2014jfm, Rigas2015jfm}.

As a macroscopic proxy for the amplitude of the symmetry-breaking, we model the evolution of the center of pressure, or the centroid of the pressure distribution on the back of the body.
The base pressure distribution is measured using 64 evenly spaced taps, as shown in Fig.~\ref{fig:wake-config}.
Since the center of pressure is a global integral quantity, we expect that fluctuations will be roughly Gaussian, based on the central limit theorem, although time correlations in the forcing still necessitates the use of subsampling and finite-time corrections.

A simple dynamical model that captures the symmetry-breaking behavior is the normal form of the pitchfork bifurcation forced by Gaussian white noise~\cite{Rigas2015jfm}.
The radial component of the Langevin equation for a symmetric two-dimensional pitchfork bifurcation forced by additive white noise is
\begin{equation}
\label{eq:cop-pitchfork}
\dot{r} = \lambda r - \mu r^3  + \frac{\sigma^2}{2r} + \sigma w(t),
\end{equation}
where the $1/r$ term appears as a consequence of Ito's lemma for a change of variables in stochastic systems. 
In one dimension, the steady-state Fokker-Planck equation can be solved analytically for this model and the free parameters can be identified based on a fit to the empirical probability distribution and mean-square displacement, as described in App.~\ref{app:ss-FP-1d}.

\begin{figure}
	\centering
	\begin{overpic}[width=0.95\linewidth]{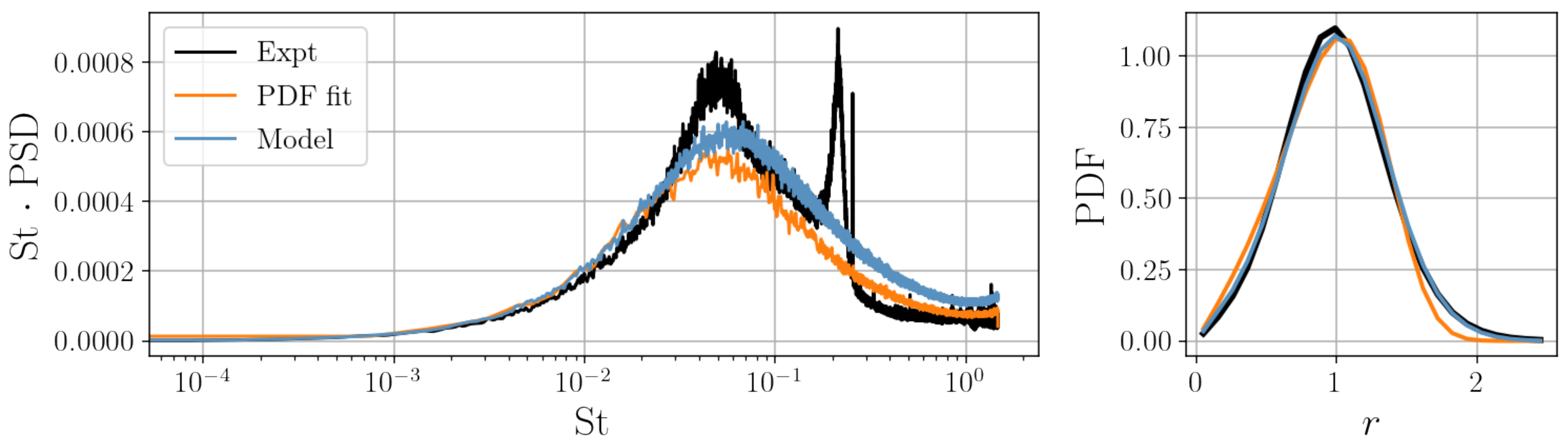}
	\end{overpic}
	\vspace{-.1in}
	\caption{ \small{ \textbf{Statistical evaluation of the axisymmetric wake models.}
			The Langevin regression model (blue) better matches both the power spectral density (left) and tails of the PDF (right) compared with the pitchfork normal form~\eqref{eq:cop-pitchfork} with coefficients estimated by PDF fitting, as in Ref.~\cite{Rigas2015jfm}.
			The models are similar, but Langevin regression identifies a quadratic state-dependent noise (Fig.~\ref{fig:wake-model}).
			The power spectrum is premultiplied by Strouhal number $\St = f U / D$, a dimensionless frequency.
			The large peak at $\St \approx 0.2$ corresponds to vortex shedding, which is essentially indistinguishable from the symmetry-breaking instability in the base pressure distribution~\cite{Brackston2016jfm}.}
	}
	\label{fig:wake-results}
\end{figure}

As shown in Fig.~\ref{fig:wake-results}, this model agrees reasonably well with the observed statistics for the center of pressure, suggesting that the stochastic modeling approach is a promising description for the leading global degrees of freedom.
However, it is difficult to extend this modeling methodology to more complex systems.
Even in one dimension, the drift and diffusion do not appear independently in the analytic steady-state PDF; additive noise can be estimated from the mean-square displacement, but this poses a challenge for multiplicative noise.
In higher dimensions, the Fokker-Planck equation does not have an analytic solution for general drift and diffusion, although model parameters might be estimated by optimizing the solution of a numerical solver.
This approach cannot resolve oscillatory behavior such as vortex shedding, since temporal information is lost in the steady-state distribution.

\begin{figure}
	\centering
	\begin{overpic}[width=0.8\linewidth]{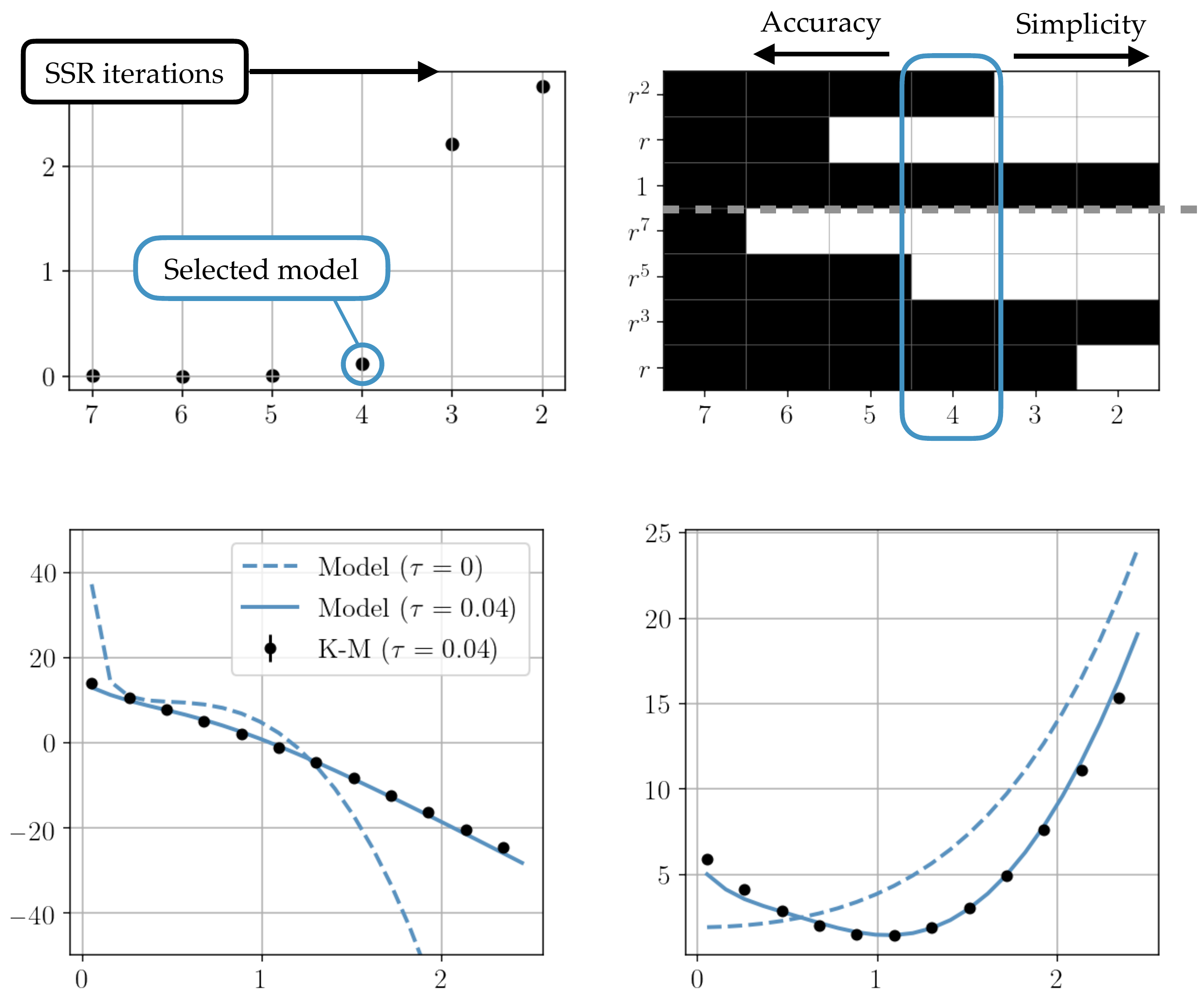}
		\small{
			\put(-1.5, 55){\begin{sideways} Cost function \end{sideways}}
			\put(15.5, 44){Number of terms}
			\put(-1.5, 16){\begin{sideways} Drift $f(r)$ \end{sideways}}
			\put(64, 44){Number of terms}
			\put(49, 13){\begin{sideways} Diffusion $a(r)$ \end{sideways}}
			\put(48.5, 54.5){\begin{sideways} Selected terms \end{sideways}}
			\put(97.5, 56){\begin{sideways} \footnotesize{Drift} \end{sideways}}
			\put(97.5, 67){\begin{sideways} \footnotesize{Diffusion} \end{sideways}}
			\put(26, -1){$r$}
			\put(76, -1){$r$}
		}
	\end{overpic}
	\caption{ 
		\vspace{0in} \small{ \textbf{Model selection and Kramers-Moyal coefficients for the axisymmetric wake.}
			The reverse-greedy Sparse Stepwise Regression identifies a hierarchy of candidate models with varying tradeoffs between accuracy and complexity.
			The optimal model has the fewest terms before the cost function begins to climb, indicating the remaining terms are essential.
			In this case, the optimal model is a pitchfork bifurcation normal form forced by quadratic multiplicative noise (top).
			The model to the right of the optimal model includes only additive noise and corresponds to the model proposed in Ref.~\cite{Rigas2015jfm}, while additional terms leads to a higher-order Stuart-Landau equation.
			The selected model closely matches the empirical finite-time Kramers-Moyal coefficients (bottom), the state PDF, and the power spectral density (Fig.~\ref{fig:wake-results}).
		}
	}
	\label{fig:wake-model}
\end{figure}
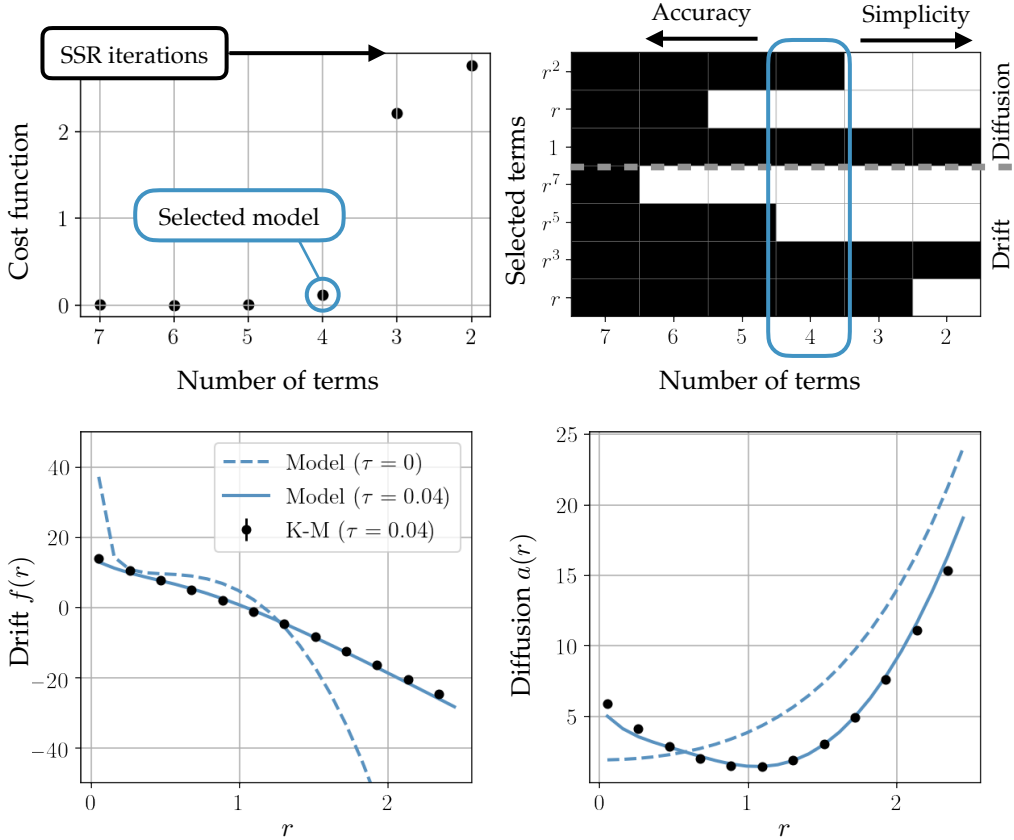

Since we do not have a known form of the model in this case, besides the intuition for the pitchfork normal form, we apply the stochastic SINDy procedure described in Sec.~\ref{sec:methods-sindy}.
Based on symmetry considerations, we include only odd polynomials in the library of drift functions.
The $1/r$ term in the drift, due to the representation in polar coordinates, is also accounted for in the optimization routine.
The model selection criteria shown in Fig.~\ref{fig:wake-model} show a clear Pareto-optimal model of the form
\begin{equation}
\dot{r} = \lambda r - \mu r^3  + \frac{\sigma^2}{2r} + (\sigma_0 + \sigma_1 r^2) w(t).
\end{equation}
Fig.~\ref{fig:wake-model} also shows that the finite-time Kramers-Moyal coefficients predicted by this model closely match those estimated by the finite-time conditional average.

The identified model is similar to that proposed by Rigas \textit{et al.}~\cite{Rigas2015jfm}, with the addition of quadratic multiplicative noise.
This modification better matches the tails of the probability distribution, as shown in Fig.~\ref{fig:wake-results}.
Monte Carlo simulation of the Langevin models also shows that the multiplicative noise leads to a more accurate power spectral density than the model based on fitting the PDF.
Quadratic multiplicative noise was previously proposed as an important modification for a spatial Langevin model of turbulence~\cite{Ragwitz2001prl}.
Multiplicative terms may be a result of neglecting degrees of freedom with bilinear coupling to the macroscopic variables~\cite{Majda2001cpam}.

This model for the symmetry-breaking instability does not resolve the peak in the power spectrum near $\St \approx 0.2$, which is related to vortex shedding in the wake~\cite{Rigas2014jfm,rigas2017weakly}.
Analysis of the corresponding laminar flow indicates that a more complete model of the wake might include three complex amplitudes to capture this periodic component and its interactions with the symmetry-breaking mode~\cite{Meliga2009jfm}.
However, the vortex shedding is only weakly observable from base pressure sensors, since it mainly takes place downstream from the body~\cite{Brackston2016jfm}; for the same reason it is less important aerodynamically than the symmetry-breaking.

Langevin regression is therefore able to clearly identify a simple low-dimensional model from experimental measurements of turbulence.
The sparse stochastic model is consistent with both known flow physics and empirical statistics, suggesting that approximating the evolution of global variables with nonlinear Langevin dynamics may be a promising direction in the low-dimensional modeling of turbulent flows.


\section{Discussion}
\label{sec:discussion}

There is a long history in the physical sciences of approximating complex multiscale systems with reduced-order models driven by stochastic forcing.
However, as this approach has spread in popularity, it is not always clear that the assumptions underpinning the rigorous treatment of nonequilibrium statistical mechanics continue to hold.
For example, the subscale degrees of freedom in systems like turbulent fluid flows often do not decorrelate fast enough to appear as delta-correlated white noise.
Nevertheless, experience suggests that simple stochastic models are often good approximations to systems that violate some of these assumptions.
Data-driven modeling has gained significant attention in this context for its ability to construct consistent empirical models without the restrictions of classical analytic approaches.

In this work, we have integrated and generalized three previously disparate approaches to data-driven stochastic modeling, combining sparse model selection based on the Kramers-Moyal average~\cite{Boninsegna2018jcp} with finite sampling-rate corrections~\cite{Lade2009, Honisch2011pre} and steady-state PDF fitting~\cite{Rigas2015jfm}.
The proposed modeling framework is designed to identify nonlinear Langevin-type dynamics from noisy experimental data, optimizing parameter estimates with both the forward and adjoint Fokker-Planck equations.
Critically, the finite-time corrections allow the method to model systems driven by correlated noise, or fast deterministic degrees of freedom, while sparse stepwise regression can be used to discover the structure of the model when it is unknown \textit{a priori}.

We have demonstrated the flexibility and generality of Langevin regression on three examples.
First, the method reconstructs a noisy Stuart-Landau oscillator from partial observations, illustrating the capability to model higher-dimensional systems.
Second, we investigated reducing the second-order dynamics of a particle in a one-dimensional double-well potential to the corresponding first-order normal form.
The data-driven model closely matches the statistical behavior of the system far from the bifurcation, well beyond the region where the analytic normal form is valid.
Finally, we apply Langevin regression to experimental measurements of a turbulent bluff body wake, identifying a model for the evolution of the base center of pressure.
In this case the SSR model selection procedure identifies a model consistent with previous work~\cite{Rigas2015jfm}, but modified with a quadratic, state-dependent noise term that better approximates the power spectrum and the long tails of the PDF.

These results indicate that the proposed method is capable of accurately modeling a broad range of systems from limited experimental observations.
However, we recognize two limitations of the method as presented here.
First, we can currently only construct first-order models.
In principle, higher-order dynamics can be recast as a system of first-order equations, provided generalized coordinates and velocities can be measured, but this is often not the case in practice.
It may therefore be easiest to model systems with stereotypical macroscopic dynamics reminiscent of a normal form or amplitude equation.
Second, although the theory generalizes naturally to higher dimensions (and we have demonstrated a two-dimensional system), the practical limitation lies in constructing $n$-dimensional histograms for Kramers-Moyal coefficients and in solving the Fokker-Planck equations on the resulting grid. 
This challenge might be addressed either with more efficient Fokker-Planck solvers or with recently-proposed methods that avoid the need for histograms at the price of enforcing consistency with the PDF~\cite{Schneider2020, Frishman2020prx, Bruckner2020prl}.

Despite these limitations, the proposed method can still be readily applied to a broad range of systems; nonlinear stochastic systems with one or two degrees of freedom have been used to model the global behavior of systems ranging from neuroscience~\cite{Laing2010neuro} to molecular dynamics~\cite{Legoll2010} and aerodynamics~\cite{Rigas2015jfm}.
The success of simple heuristics combined with stochastic models for feedback control (e.g.~\cite{Brackston2016jfm}) also suggests that these data-driven models could be integrated in a control scheme, although this will require modeling the effects of actuation on the system in addition to the homogeneous behavior.

Another goal of modeling is to uncover the latent low-dimensional structure of macroscopic dynamics.
In fluid dynamics, for instance, this is often conceptualized as a small set of global modes whose amplitudes evolve according to low-dimensional nonlinear dynamics~\cite{Holmes1996}.
Moving beyond integral quantities such as the center of pressure, Langevin regression could be used to identify a data-driven, stochastic counterpart to Galerkin-type reduced-order models and resolve important nonlinear interactions between the large-scale structures of the flow.

In a broader context, the ability to identify reduced-order stochastic models from noisy experimental measurements of multiscale nonlinear dynamics will unlock a powerful set of tools for a much wider range of systems.
Data-driven methods allow for the treatment of not only systems which break the strict assumptions of classical stochastic modeling, but also data from ecology, epidemiology, and neuroscience for which first-principles governing equations are unavailable.
Tools such as the Kramers-Moyal average already have a history of success in a variety of fields. 
It is our hope that the proposed method will build on this legacy and extend these successes to an even more extensive class of complex systems.

\section*{Acknowledgments} JLC acknowledges support from the NDSEG fellowship.  The authors acknowledge funding support from the Army Research Office (ARO W911NF-19-1-0045) and the Air Force Office of Scientific Research (AFOSR FA9550-18-1-0200).  We would also like to thank Nathan Kutz, Cecilia Clementi, Lionel Mathelin, Edouard Boujo, and Bernd Noack for valuable discussions about nonlinear modeling and stochastic systems.  

\appendix

\section{Determination of sampling rate}
\label{app:sampling-rate}

\begin{SCfigure}
	\centering
	\begin{overpic}[width=0.6\linewidth]{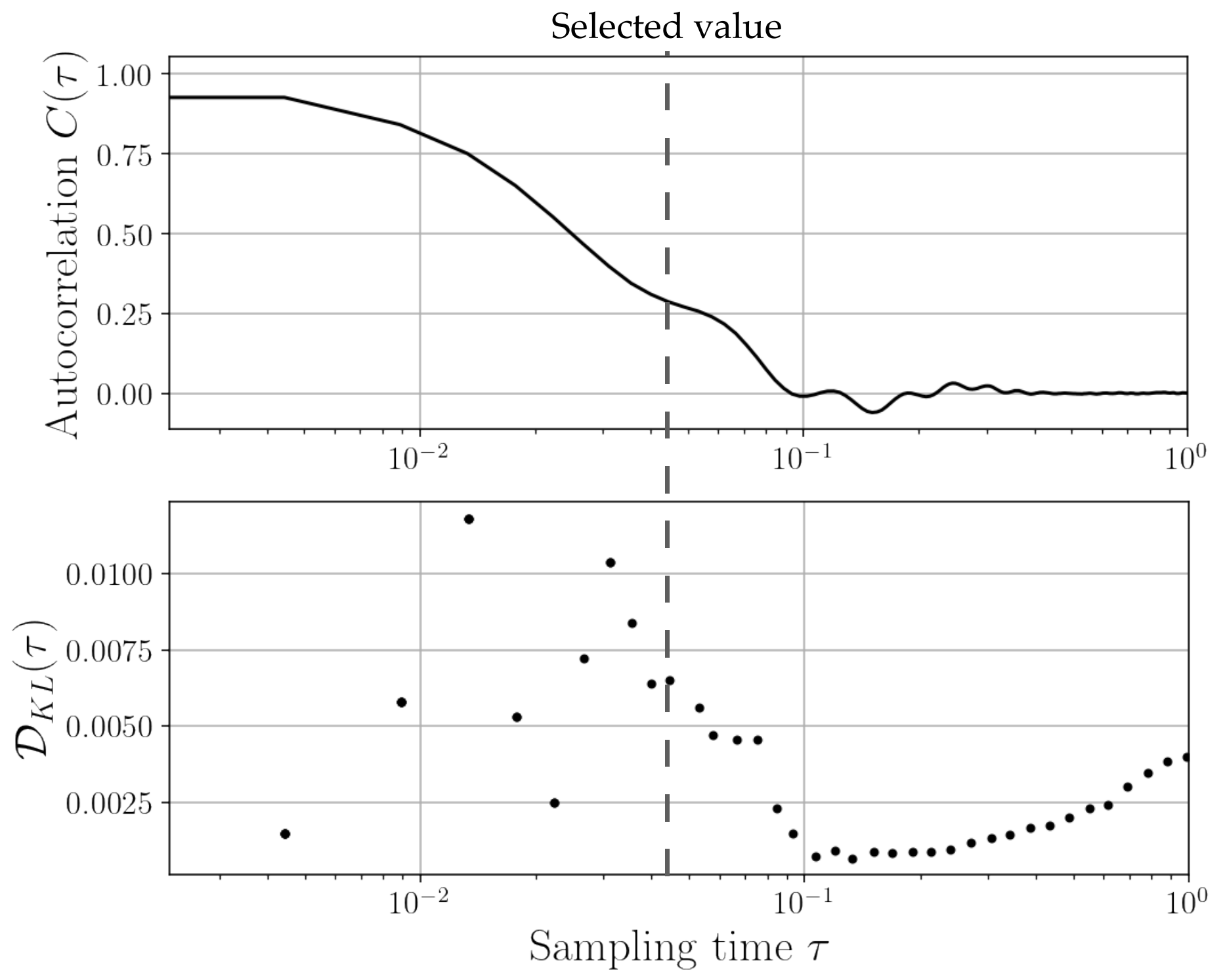}
	\end{overpic}
	\caption{ 
		\small{\textbf{Diagnostic statistics for finite-time sampling rate.}
	    The coarse sampling rate allows the fast fluctuations to decorrelate while still resolving the macroscopic dynamics.
	    For a good choice of sampling rate, the autocorrelation function will take on an intermediate value and the K-L divergence between the three-time PDF and its Markov approximation will be near a minimum.
	    As seen for the turbulent wake data, an appropriate value may need to balance these two requirements.
		}
		\vspace{0.3in}
	}
	\label{fig:sampling-rates}
\end{SCfigure}

As discussed above, the choice of sampling rate for the Kramers-Moyal average is critical, and the experimental sampling frequency is not necessarily the best choice.
For systems driven by Gaussian white noise, the fastest sampling rate possible yields the best approximation to the true zero-time Kramers-Moyal coefficients.
However, sampling multiscale systems with broadband spectral content too quickly can lead to significantly underestimating the conditional moments due to time correlations in the ``noise".
A rigorous condition for the optimal sampling rate in general is still unknown; however, this appendix presents several complementary diagnostic tools.
These methods are discussed for scalar variables $x(t)$, but similar conclusions hold for high-dimensional systems.
Fig.~\ref{fig:sampling-rates} illustrates the sampling rate determination for the turbulent wake data.

\subsection{Autocorrelation time}
In view of the dual scale separation principle illustrated in Fig.~\ref{fig:scale-separation}, the determination of sampling rate is essentially a question of the dominant time scales in the data.
One of the simplest and most revealing statistics is the autocorrelation function, defined as
\begin{equation}
    C(\tau) = \frac{ \langle x(t+\tau) x(t) \rangle_t } {\langle x(t)^2 \rangle}.
\end{equation}

By construction, $C(0) = 1$, and for most complex systems $C(\tau)$ approaches zero on some characteristic time scale.
However, this time scale is generally related to the macroscopic dynamics; it is not desirable to sample the system once the macroscopic dynamics have completely decorrelated.
On the other hand, if the data is sampled while $C(\tau) \sim 1$, the correlations will tend to eclipse the effects of the unresolved scales (Fig.~\ref{fig:doublewell-km} top, open circles).
Intermediate values of the autocorrelation function (e.g. 0.2 to 0.8) indicate time scales where the fast scales introduce significant fluctuations, but the macroscopic dynamics have not yet fully decorrelated.

Similar intuition can be gained from the empirical power spectral density, which is related to the autocorrelation function according to the Wiener-Khinchin theorem.
Peaks in the power spectrum are typically related to the macroscopic dynamics; the Nyquist frequency then indicates an upper bound on the sampling rate.

\subsection{Markov test}

A more rigorous procedure for determining sampling intervals was introduced in Ref.~\cite{Friedrich1998epl} based on testing the Markov property of the system at different sampling rates.
Markovian dynamics depend only on the state of the system at the current time.
This implies that the conditional PDF for the evolution of the system does not depend on any earlier times:
\begin{equation}
    p(x_3, t + \tau | x_2, t; x_1, t-\tau) = p(x_3, t + \tau | x_2, t).
\end{equation}
In fact, this assumption is central to the derivation of the Fokker-Planck equation~\cite{Risken1996book}.

Using the definition of conditional probability, the Markov property implies that
\begin{equation}
    p(x_3, t + \tau ; x_2, t; x_1, t-\tau) = p(x_3, t + \tau | x_2, t) p(x_2, t; x_1, t-\tau).
\end{equation}
This can be directly tested by forming the left and right sides and comparing them for equality for various sampling rates $\tau$.
Earlier work has used a least-squares comparison of the PDFs, but we instead evaluate the Kullback-Leibler divergence $\mathcal{D}_{KL}(\tau)$ between the three-time joint PDF and the Markov approximation for consistency with the K-L divergence in the optimization problem.

At short times, ignoring the fast time scales leads to non-Markovian effects in the macroscopic variables.
Eventually, if $\tau$ is large enough that the fluctuations can decorrelate, the K-L divergence reaches a minimum.
In practice, we seek to balance this condition with considerations of the autocorrelation function.
We therefore choose a sampling rate for which $C(\tau)$ takes on intermediate values and $\mathcal{D}_{KL}(\tau)$ is at or approaching its minimum.

\section{Steady-state Fokker-Planck solvers}
\label{app:ss-FP}

This section describes three numerical solvers for the steady-state Fokker-Planck equation.
The forward Fokker-Planck equation governs the time evolution of the PDF $p(x)$, where in general $x$ may be a $d-$dimensional state vector.
We assume the state evolves according to a Langevin equation
\begin{equation}
    \dot{x}_i = f_i(x, t) + \sigma_i(x, t) w_i(t),
\end{equation}
where each $w_i(t)$ is an independent Gaussian white noise (Wiener) process.

If the drift and diffusion functions do not depend on time, i.e. $f_i(x, t) = f_i(x)$ and $\sigma_i(x, t) = \sigma_i(x)$, the PDF $p(x)$ is a solution to the steady-state Fokker-Planck equation
\begin{equation} \label{eq:fokker-planck-ndim}
0 = - \pdv{}{x_i} f_i(x) p(x) + \frac{\partial^2}{\partial x_i \partial x_j} a_{ij}(x) p(x),
\end{equation}
where $a_{ij}(x) = \frac{1}{2} \sigma_i(x) \sigma_j(x)$ is the diffusion tensor.
For the sake of simplicity, we will restrict the discussion to the common case of diagonal diffusion, so that $a_{ij} = 0$ when $i \neq j$.

The steady-state PDF $p(x)$ is subject to the normalization condition
\begin{equation}\label{eq:pdf-norm}
1 = \int_{\R^d} p(x) \dd x.
\end{equation}

\subsection{Exact solution in one dimension}
\label{app:ss-FP-1d}

In one dimension, the steady-state Fokker-Planck equation~\eqref{eq:fokker-planck-ndim} can be integrated explicitly.
For additive noise $\sigma$, the solution has a potential-like form:
\begin{subequations}
\begin{align}
    p(x) &= C \exp{-\frac{2}{\sigma^2} U(x)}\\
    U(x) &= -\int f(x) \dd x,
\end{align}
\end{subequations}
where the constant $C$ is given by the normalization condition~\eqref{eq:pdf-norm}.
For state-dependent noise, the solution has the form
\begin{equation}
\label{eq:app-1d-ss-fp}
p(x) = \frac{C}{a(x)}\exp \left[ \int \frac{f(x)}{a(x)} \dd x  \right].
\end{equation}
In either case, the ``potential" and normalization integrals can be evaluated numerically for given drift and diffusion functions.

In principle, this solution can be used to fit model parameters against the empirical PDF.
However, only the ratio $f(x) / a(x)$ appears in the solution, so that an independent estimate of the noise amplitude $\sigma$ is necessary.
For constant diffusion, this can be estimated from the mean-square displacement of the radial coordinate, which is predicted to grow as $ \langle (\Delta x(\tau)) ^2 \rangle = \sigma^2 \tau $ for short times.

This method, labeled as ``PDF fit" in Fig.~\ref{fig:wake-results} works well for simple scalar Langevin equations with additive white noise, but does not generalize to more complicated models.
A numerical approximation to the analytic solution~\eqref{eq:app-1d-ss-fp} can still be used in the Langevin regression optimization problem to enforce consistency with the steady-state PDF.

\subsection{Fourier-Galerkin solver for steady-state Fokker-Planck equation}
This section describes a steady-state solver based on~\cite{Soize1988} where we approximate the PDF with a Fourier series and then Galerkin project the equation onto the Fourier modes.

We use the Fourier representation of the pdf:
\begin{subequations}
\begin{gather}
p(x) = \frac{1}{2\pi} \int_{-\infty}^\infty \hat{p}(k) e^{ikx} \dd k \\
\hat{p}(k) = \int_{-\infty}^\infty p(x) e^{-ikx} \dd x.
\end{gather}
\end{subequations}
In this case the normalization condition implies $\hat{p}(0) = 1$.

\paragraph{One-dimensional formulation.}

In practice, the Fourier representation will be truncated at a finite number of modes.  After substituting the approximate representation of $p(x)$ we find the residual
\begin{equation}
R = \frac{1}{2\pi} \int \hat{p}(k)  \dd k \left[ -\pdv{}{x} \left(f(x) e^{ikx} \right) + \pdv[2]{}{x} \left(a(x) e^{ikx}\right) \right]. 
\end{equation}
Minimizing the residual requires that it is orthogonal to the subspace spanned by the Fourier modes.
Therefore we project onto an arbitrary wavenumber $k'$:
\begin{subequations}
\begin{align}
0 &= \int \hat{p}(k) \dd k \int e^{-ikx} \dd k'  \left[ -\pdv{}{x} \left(f(x) e^{ikx} \right) + \pdv[2]{}{x} \left(a(x) e^{ikx}\right) \right]  \\
&= \iint \hat{p}(k) \dd k \dd k' \int e^{-i(k'-k)x}  \left[ -ik' f(x) - {k'}^2 a(x) \right] \\
\Longrightarrow\quad 0 &= \int  \hat{p}(k) \dd k \left[ -ik' \hat{f}(k'-k) - {k'}^2 \hat{a}(k'-k) \right].
\end{align}
\end{subequations}
This equation must be true for all $k'$.  Practically, the FFT is used, so the integral is a sum over wavenumbers.  Similar to the Hermite representation, we can use the normalization condition to obtain an inhomogeneous linear equation:
\begin{subequations}
\begin{gather}
b(k') = ik' \hat{f}(k') + {k'}^2 \hat{a}(k') \\
A (k', k) = -ik' \hat{f}(k'-k) - {k'}^2 \hat{a}(k'-k) \\
b(k') = \sum_{k \neq 0} A(k', k) \hat{p}(k) .
\end{gather}
\end{subequations}
Then we simply invert $\hat{p}(k)$ to obtain the PDF.

\paragraph{Generalization to $d$ dimensions.}

In higher dimensions $x \in \R^d$, the steady-state Fokker-Planck equation is
\begin{equation}
0 = -\pdv{x_n}f_n(x) p(x) + \frac{\partial^2}{x_n x_m} a_{nm}(x) p(x) .
\end{equation}
The procedure is similar, but the linear equation now involves vector wavenumbers $k, k' \in \R^d$:
\begin{subequations}
\begin{gather}
b(k') = ik_n' \hat{f}_n(k') + k_n' k_m' \hat{a}_{nm}(k') \\
A (k', k) = -ik_n' \hat{f}_n(k'-k) - k_n' k_m' \hat{a}_{nm}(k'-k) \\
b(k') = \sum_{k \neq 0} A(k', k) \hat{p}(k) .
\end{gather}
\end{subequations}

\subsection{Pseudospectral Arnoldi method}

Constructing the Galerkin operators becomes increasingly difficult in higher dimensions; an alternative is to avoid explicitly constructing the Fokker-Planck operators and instead approximate the steady-state solution with a time-stepping method~\cite{Antoulas2005}.
Assuming the Fokker-Planck equation has a unique steady-state solution, this solution is the eigenvector of the Fokker-Planck operator $\mathcal{L}$ with zero eigenvalue.
This eigenvector can be approximated with matrix-free time-stepping methods, such as Arnoldi iteration.
Efficient implementations of the implicitly restarted Arnoldi method are available in many programming languages, such as Python (SciPy), Julia (IterativeSolvers.jl), and Fortran (ARPACK).
This approach only requires defining the action of the operator $\mathcal{L}$ on a probability distribution $p$ (a ``matrix-vector" product).
We discretize the Fokker-Planck operator with a pseudospectral approach:
\begin{equation}
    \mathcal{L}p = \sum_{n=0}^d \mathcal{F}^{-1}_n \left\{ -ik_n\mathcal{F}_n\{f_n(x)p(x)\} - k_n^2 \mathcal{F}_n\{a_n(x)p(x)\} \right\},
\end{equation}
where $\mathcal{F}_n$ and $\mathcal{F}^{-1}_n$ denote the forward and inverse Fourier transforms in the $n$-th dimension.

\section{Normal form approximation to the second-order dynamics}
\label{app:normal-form}

We model a particle of unit mass in a symmetric double-well potential subject to thermal fluctuations with the second-order Langevin dynamics
\begin{equation}
    \ddot{x} + 2 \gamma \dot{x} = \alpha x - \beta x^3 + \sqrt{2 \gamma k_\mathrm{B} T} w(t).
\end{equation}
Nondimensionalizing with the relaxation timescale $\gamma^{-1}$ and the length scale $\sqrt{\gamma^2/\beta}$ given by the nonlinear term, the dynamics are
\begin{equation}
    \ddot{x} + 2 \dot{x} = \epsilon x - x^3 + \sigma_x w(t),
\end{equation} 
where $\epsilon = \alpha/\gamma^2$ and $\sigma_x^2/2$ is the dimensionless energy of the thermal fluctuations.

The system linearized at the origin has eigenvalues $\lambda_{1, 2} = 1 \pm \sqrt{1 + \epsilon} $, indicating the system undergoes a supercritical pitchfork bifurcation at $\epsilon_c = 0$. 
The state of the system is given by
\begin{equation}
\begin{bmatrix}
x \\ \dot{x}
\end{bmatrix} = v_1 \phi_1(t) + v_2 \phi_2(t),
\end{equation}
where $v_{1, 2}$ are the corresponding eigenvectors. 
Close to the bifurcation, we assume the following:
\begin{enumerate} \setlength{\itemsep}{-3pt}
    \item \textbf{Invariant manifold reduction:} 
    The dynamics are restricted to the one-dimensional subspace spanned by the unstable eigenvector $v_1$.
    Since the eigenvectors are not orthogonal, the amplitude of the stable eigenvector is an algebraic function of $\phi_1(t)$ on restriction to this subspace.
    To leading order, $\phi_2 = h\phi_1$.
    \item \textbf{Normal form:} The drift dynamics in this subspace are given by the normal form for the pitchfork bifurcation with unknown parameter $\mu$:
    \begin{equation}
        \dot{\phi}_1 = \lambda_1(\epsilon) \phi_1 -\mu(\epsilon) \phi_1^3 + \sigma_\phi w(t).
    \end{equation}
    \item \textbf{Dynamical consistency:} The reduced-order dynamics preserve the fixed points of the drift function of the full system.  That is, $\ddot{x} = \dot{x} = 0$ at $x = 0, \pm \sqrt{\epsilon}$.
\end{enumerate}
These assumptions imply that $h = -\lambda_1/\lambda_2$ and $\mu = (1+h)^2 \lambda_1/\epsilon$.
Furthermore, under the similarity transform and It\^o's lemma for a change of variables~\cite{Risken1996book} the diffusion coefficient becomes~$\sigma_\phi = \sigma_x / 2 \sqrt{1 + \epsilon}$.
Inverting the transformation, $x \approx (1+h)\phi_1$, which can be used to recast the eigenvector dynamics into the form of Eq.~\eqref{eq:doublewell-nf}.
	
\newpage
\begin{spacing}{.01}
	\small
	\bibliographystyle{unsrt}

\end{spacing}
\end{document}